\documentclass[preprint,prb]{revtex4-1}
\usepackage{geometry}
\usepackage[section]{placeins} 
\usepackage{dblfloatfix} 

\usepackage{bm}
\usepackage{amsmath,amssymb}
\usepackage{graphicx}
\usepackage[hidelinks,colorlinks=true,linkcolor=black,citecolor=blue]{hyperref}
\usepackage[nameinlink]{cleveref}
\usepackage[labelsep=period,justification=raggedright,singlelinecheck=false]{caption}

\newcommand*{\hamiltonian}{\mathcal{H}}
\newcommand*{\transpose}[1]{#1^\mathrm{T}}
\newcommand*{\inverse}[1]{#1^{-1}}
\newcommand*{\sumnear}{\sum_{\langle i,j \rangle}^{\phantom{1}}}
\newcommand*{\trace}{\mathrm{Tr}\,}
\newcommand*{\ii}{\mathrm{i}}
\newcommand*{\dd}{\mathrm{d}}
\newcommand*{\sg}{\mathrm{SG}}
\newcommand*{\fm}{\mathrm{FM}}
\newcommand*{\Tc}{T_\mathrm{c}}
\newcommand*{\ypredict}{Y_{\mathrm{predict}}}
\newcommand*{\thetaOpt}{\theta_{\mathrm{opt}}}
\newcommand*{\Set}[1]{\left| #1 \right|}
\newcommand*{\ConfigAve}[1]{\left[ #1 \right]_{\mathrm{c}}}
\newcommand*{\configAve}[1]{\Big[ #1 \Big]_{\mathrm{c}}}

\makeatletter
\renewcommand{\p@subsection}{}
\renewcommand{\p@subsubsection}{}
\makeatother

\crefname{section}{Sect.}{Sects.}
\Crefname{section}{Section}{Sections}
\crefname{equation}{Eq.}{Eqs.}
\Crefname{equation}{Equation}{Equations}
\crefname{figure}{Fig.}{Figs.}
\Crefname{figure}{Figure}{Figures}
\crefname{table}{Table}{Tables}
\Crefname{table}{Table}{Tables}

\begin{document}

\author{Yosei Takada}
\affiliation{Department of Engineering Science, Graduate School of Informatics and Engineering, The University of Electro-Communications, 1-5-1 Chofugaoka, Chofu-shi, Tokyo 182-8585, Japan}
\author{Yusuke Terasawa}
\affiliation{Department of Engineering Science, Graduate School of Informatics and Engineering, The University of Electro-Communications, 1-5-1 Chofugaoka, Chofu-shi, Tokyo 182-8585, Japan}
\author{Yuma Osada}
\affiliation{Department of Engineering Science, Graduate School of Informatics and Engineering, The University of Electro-Communications, 1-5-1 Chofugaoka, Chofu-shi, Tokyo 182-8585, Japan}
\author{Yukiyasu Ozeki}
\affiliation{Department of Engineering Science, Graduate School of Informatics and Engineering, The University of Electro-Communications, 1-5-1 Chofugaoka, Chofu-shi, Tokyo 182-8585, Japan}

\title{Critical Universality of the $SU(2)$ Gauge Glass Model Analyzed by the Dynamical Scaling Method}
\date{\today}
\begin{abstract}
We investigate the critical phenomena of the three-dimensional (3D) $SU(2)$ gauge glass model, which can be regarded as the $O(4)$ spin-glass model with gauge symmetry.
Using Monte Carlo simulations and the non-equilibrium relaxation method, we examine the critical behavior along phase boundaries across various degrees of disorder.
Consistent with previous studies, we verify that weak disorder is irrelevant to the universality class; the critical behavior of the ferromagnetic-paramagnetic transition remains in the 3D $O(4)$ universality class of the pure ferromagnetic system.
Additionally, we demonstrate that the paramagnetic-spin glass transition exhibits universal critical behavior independent of the disorder.
Our findings provide valuable insights into the fundamental properties and universality of gauge glass models.
\end{abstract}

\maketitle

\section{Introduction}

    The physics of disordered systems, particularly that of spin glasses (SGs), has been a subject of long-standing interest.~\cite{1975JPhF5965E,PhysRevLett.35.1792,RevModPhys.58.801}
    Because of their complex interactions, frustration, and slow dynamics, the numerical analysis of SGs remains challenging. Therefore, extensive research has been conducted to elucidate their phase diagrams and fundamental properties.~\cite{Charbonneau2023}

    In this context, gauge theory~\cite{10.1143/PTP.66.1169, Ozeki1993} has been established as an analytical framework for studying SGs by exploiting gauge symmetry. Models possessing such symmetry are referred to as gauge glass (GG) models.
    In GG models, several exact relations can be derived analytically within the subspace known as the Nishimori line (NL) on the phase diagram. Furthermore, this line imposes strong constraints on the overall structure of the phase diagram.
    Typical examples of GGs include the Ising SG and the XY SG with random phase shifts (hereafter referred to as the XY GG).
    These models have been extensively studied because they are related to error-correcting codes, NP-hard problems, and Josephson junction arrays.~\cite{Sourlas1989, F_Barahona_1982, PhysRevB.33.6533, PhysRevB.42.1059}
    In recent years, motivated by the early observation that quenched disorder in real spin-glass materials exhibits spatial correlations,~\cite{Mydosh1993} an increasing number of studies have employed gauge theories to rigorously understand these properties.~\cite{PhysRevE.110.064108, PhysRevE.111.044109, qp1w-qcbs}

    Although the framework of gauge theory can be applied to various SG models~\cite{Ozeki1993}, previous studies on GGs have mainly focused on the Ising and XY models.
    The $SU(2)$ GG model with local $SU(2)$ symmetry has been proposed, but its physical properties are still not well understood.
    This historical focus has hindered a comprehensive discussion on the universality classes of GG models.
    First, continuous-spin GG models to which gauge theory can be applied are currently limited to the XY and $SU(2)$ cases.
    By analyzing the latter, we aim to clarify the common properties of continuous-spin GG models.
    Second, at the multicritical point (MCP) on the NL, the possibility of superuniversality has been suggested.~\cite{Delfino_2025}
    However, numerical verification of this hypothesis has so far been limited primarily to the Ising SG and the XY GG.
    By investigating the $SU(2)$ GG model, which has different spin degrees of freedom, we expect to gain new insights into this prediction.

    In this study, we numerically investigate the critical behavior of the ferromagnetic-paramagnetic (FM-PM) and PM-SG transitions in the 3D $SU(2)$ GG model by tuning the degrees of disorder from the pure-system limit.
    For the FM-PM transition, the Harris criterion~\cite{A_B_Harris_1974} indicates that if the pure system has the specific-heat exponent $\alpha < 0$, quenched disorder is irrelevant to the universality class, whereas if $\alpha > 0$, it becomes relevant.
    A previous study has shown that the $O(4)$ model, which corresponds to the pure limit of the $SU(2)$ GG model, has
    $\alpha = -0.244(27)$.~\cite{PhysRevD.51.2404}
    Following this criterion, we expect the system to belong to the same universality class as the $O(4)$ model at weak disorder.
    Therefore, we vary the disorder parameter to confirm this universality class and verify whether the critical exponents at the MCP exhibit superuniversality.
    For the PM-SG transition, we demonstrate the existence of a finite-temperature SG phase in the $SU(2)$ GG model.
    Furthermore, we verify whether the disorder-independent universality of the critical exponents, as reported in other SG models,~\cite{doi:10.7566/JPSJ.92.074003,PhysRevB.83.094203} also holds in our model.

    To carry out the analysis described above, we apply the non-equilibrium relaxation (NER) method~\cite{ozeki2007nonequilibrium} based on Monte Carlo simulations.
    We perform dynamical scaling to estimate the transition temperature, utilizing magnetization relaxation for the FM transition and SG susceptibility for the SG transition.
    The introduction of Gaussian process regression (GPR) has streamlined the dynamical scaling analysis procedure and dramatically improved the accuracy of the results.~\cite{PhysRevE.94.043312}
    Although dynamical scaling analysis allows for the simultaneous evaluation of several critical exponents, these are influenced by the finite observation time and thus exhibit some deviation from their thermodynamic values.
    Such effects do not appear significantly in the evaluation of transition temperatures and are therefore reliable.
    Thus, we re-simulate the relaxation of several fluctuating physical quantities, including the order parameter, precisely at the obtained transition temperature, and evaluate the thermodynamic critical exponent with high precision by determining its asymptotic power law.
    This requires high-precision numerical differentiation using GPR and the extrapolation of the obtained power exponents to infinite time, utilizing methods applied to FM transitions and SG transitions.~\cite{doi:10.7566/JPSJ.93.114001}

    It is noted that when analyzing SG transitions using the NER method---particularly when applied to continuous spin systems---complex relaxation processes often arise, making the analysis difficult.~\cite{terasawa2026sg}
    This is thought to be largely due to the influence of chiral glass transitions that occur in continuous spin systems, and the fact that their relationship with SG transitions remains unresolved is considered a major factor.
    On the other hand, in the GG model, chiral glass transitions are generally not expected to occur because the mirror symmetry in spin space is broken in the Hamiltonian.
    Consequently, as demonstrated in this study, SG analysis using the NER method in the GG model is sufficiently stable, making it an excellent example of the method's application to SG systems.

    The remainder of this paper is organized as follows.
    In \cref{sec:Models}, we define the model and observables.
    \cref{sec:method} describes the NER method used to estimate the critical temperature and critical exponents.
    In \cref{sec:result}, we present the simulation results obtained using this method.
    Finally, in \cref{sec:discussion}, we provide our discussions and future perspectives.

\section{Models}
\label{sec:Models}

\subsection{Gauge glass model}
\label{subsec:gauge_glass}

\begin{figure}[htbp]
        \centering
        \includegraphics[width=6cm]{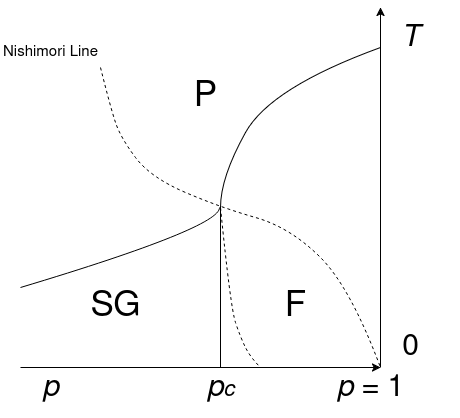}
        \caption{Phase diagram of the $\pm J$ Ising SG model, which is one of the GG models.
        F, P, and SG denote the FM, PM, and SG phases, respectively.
        The case of $p=1$ corresponds to the pure system without disorder.}
    \label{figure:spin_glass_phase_diagram}
    \end{figure}

    SG models that satisfy local gauge symmetry and allow the application of gauge theory are called GG models.
    A typical phase diagram of such models is shown in \cref{figure:spin_glass_phase_diagram}.
    Within this phase diagram, a special subspace known as the NL is defined by a specific relation between the temperature and the disorder parameter.
    On this line, exact expressions for the internal energy, an upper bound for the specific heat, and exact identities for correlation functions can be derived regardless of the spatial dimension or lattice structure.
    Furthermore, the NL imposes strict constraints on the phase diagram.
    For example, the NL passes through the MCP but never enters the SG phase, and the phase boundary between the FM and SG phases below the MCP must be vertical or reentrant.

    Examples of GG models include the $\pm J$ and Gaussian Ising SG models, the XY GG model, and random Potts models with $Z_q$ symmetric bonds.

\subsection{O(4) model (pure system)}
\label{subsec:o4_model}

    First, let us consider the $SU(2)$ GG model in the absence of quenched disorder. The Hamiltonian is given by
\begin{equation}
    \hamiltonian \{ \phi \} = -J \sumnear \trace (\phi_i \phi_j^{\dagger}), \label{eq:o4_hamiltonian}
\end{equation}
where $J> 0$ is the interaction constant, $\sum_{\langle i,j \rangle}$ denotes the sum over nearest-neighbor pairs, and $\phi_i$ is the spin variable represented by a $2 \times 2$ non-Abelian special unitary matrix in $SU(2) = \{ g \in U(2) \mid \, \mathrm{det} \, g = 1\}$.
By introducing four real numbers $\sigma_i$ and $\bm{\pi}_i = (\pi_{i,1}, \pi_{i,2}, \pi_{i,3})$, the spin variable $\phi_i$ can be parameterized as
\begin{equation}
    \phi_i = \sigma_i I  + \ii \bm{\tau} \cdot \bm{\pi}_i,
\end{equation}
where $I$ is the identity matrix and $\bm{\tau} = (\tau_1, \tau_2, \tau_3)$ represents the Pauli matrices.
Expanding $\phi_i$ yields
\begin{equation}
    \phi_i =
    \begin{pmatrix}
       \sigma_i + \ii \pi_{i,3} & \pi_{i,2} + \ii \pi_{i,1} \\
       -\pi_{i,2} + \ii \pi_{i,1} & \sigma_i - \ii \pi_{i,3}
    \end{pmatrix}.
\end{equation}
The $SU(2)$ group condition, $\mathrm{det}\, \phi_i = 1$, imposes the following geometric constraint on the variable:
    \begin{equation}
        \mathrm{det} \, \phi_i = \sigma_i^2 + \pi_{i,1}^2 + \pi_{i,2}^2 + \pi_{i,3}^2 = 1.
    \end{equation}
Furthermore, by expanding the Hamiltonian in \cref{eq:o4_hamiltonian}, we obtain
    \begin{equation}
        \label{eq:o4_expansion}
        \begin{aligned}
            \hamiltonian\{ \phi \} &= -J \sumnear \trace (\phi_i \phi_j^{\dagger}) \\
                         &= -2J \sumnear (\sigma_i \sigma_j + \bm{\pi}_i \cdot \bm{\pi}_j) \\
                         &= -2J \sumnear \bm{S}_i \cdot \bm{S}_j,
        \end{aligned}
    \end{equation}
    where $\bm{S}_i = (\sigma_i, \pi_{i,1},\pi_{i,2},\pi_{i,3})$ and $|\bm{S}_i| =  1$.
    This result shows that the Hamiltonian is equivalent to the $O(4)$ model, which is described by four-component unit continuous spins.
    The invariant measure for the spin variables, $\dd \mu\{ \phi \}$, is given by
    \begin{equation}
        \label{eq:spin_measure}
        \dd \mu(\phi_i) = \delta(\sigma_i^2 + \bm{\pi}_i \cdot \bm{\pi}_i - 1) \dd \sigma_i \dd \bm{\pi}_i,
    \end{equation}
    where $\delta$ is the Dirac delta function. Using the expressions derived above, the thermal average of an arbitrary observable $\mathcal{O}\{\phi\}$ is expressed as
    \begin{equation}
        \langle \mathcal{O} \rangle = Z(\beta)^{-1} \int \dd \mu\{\phi\} \mathcal{O}\{\phi\} \exp(- \beta \hamiltonian \{ \phi \}),
    \end{equation}
    where $\beta = 1/k_\mathrm{B} T$ is the inverse temperature, and $T$ is the temperature. The partition function is defined as
    \begin{equation}
        Z(\beta) = \int \dd \mu\{\phi\} \exp(- \beta \hamiltonian \{ \phi \}).
    \end{equation}
    According to a previous study,~\cite{PhysRevD.29.338} if the finite-temperature chiral phase transition in two-flavor massless QCD is of second order, its universality class is expected to belong to the 3D $O(4)$ model.

\subsection{$SU(2)$ gauge glass model}
\label{subsec:su2gg_model}

    In the following, let us consider the case where quenched disorder is introduced.
    The Hamiltonian of the $SU(2)$ GG model is given by
    \begin{equation}
        \label{eq:su2_hamiltonian}
        \hamiltonian \{ \phi \} \{ \omega \} = -J \sumnear \trace (\phi_i \omega_{ij} \phi_j^{\dagger}) ,
    \end{equation}
    where $\omega_{ij} \in SU(2)$ is a quenched random variable that can be parameterized as $\omega_{ij} = x_{ij} + \mathrm{i} \bm{\tau} \cdot \bm{y}_{ij}$.
    We consider the following distribution of $\omega_{ij}$:
    \begin{equation}
        \label{eq:su2_probability_density}
        P(\omega_{ij}) = \frac{D}{2 \pi^2 I_1(2D)} \exp(D \,  \trace  \omega_{ij}),
    \end{equation}
    where $D$ is the disorder parameter and $I_n(x)$ is the modified Bessel function.
    Note that $\trace \omega_{ij} = 2x_{ij}$.
    In the limit $1/D \to 0$, we have $\omega_{ij} = I$ ($x_{ij} = 1$ and $y_{ij} = 0$), and the Hamiltonian of the $SU(2)$ GG model reduces to that of the $O(4)$ model.
    Conversely, in the limit $1/D \to \infty$, $\omega_{ij}$ obeys the uniform distribution.

    The invariant measure for the quenched random variables, $\dd \nu \{ \omega \}$, is derived in the same manner as for the spin variable, yielding
    \begin{equation}
        \label{eq:measure_bond}
        \dd \nu(\omega_{ij}) = \delta(x_{ij}^2 + \bm{y}_{ij} \cdot \bm{y}_{ij} - 1) \dd x_{ij} \dd \bm{y}_{ij}.
    \end{equation}
    Analogously, the configuration average can be defined as:
    \begin{equation}
        \ConfigAve{\mathcal{O}} = \int \dd \nu\{\omega\} \mathcal{O}\{ \omega \} P\{ \omega \}.
    \end{equation}
    To demonstrate the gauge invariance of the system, we introduce the local gauge variable $\theta_i \in SU(2)$ and define the gauge transformation as follows:
    \begin{equation}
    \begin{aligned}
        \phi_i &\to \phi_i \, \theta_i,\\
        \omega_{ij} &\to \theta_i^{-1} \omega_{ij} \, \theta_j.
    \end{aligned}
    \end{equation}
    The Hamiltonian satisfies the following invariance:
    \begin{equation}
        \label{eq:gauge_transformation}
        \hamiltonian \to -J \sumnear \trace (\phi_i \theta_i \theta_i^{-1} \omega_{ij} \theta_j \theta^{-1}_j \phi^{\dagger}_j) = \hamiltonian,
    \end{equation}
    where we have used the relation $A^{-1} = A^{\dagger}$ for a unitary matrix $A$.
    According to \cref{eq:gauge_transformation}, we find that \cref{eq:su2_hamiltonian} with \cref{eq:su2_probability_density} satisfies the gauge symmetry.

    \Cref{eq:su2_hamiltonian} can be expanded as follows:
    \begin{equation}
        \label{eq:expansion}
    \begin{aligned}
        \hamiltonian = -2J \sumnear &\left\{ (\sigma_i \sigma_j + \bm{\pi}_i \cdot \bm{\pi}_j ) \,x_{ij} \right.\\
        &  + \left. (\bm{\pi}_i \times \bm{\pi}_j + \sigma_i\bm{\pi}_j - \sigma_j\bm{\pi}_i ) \cdot \bm{y}_{ij} \right\}.
    \end{aligned}
    \end{equation}
    The Hamiltonian corresponds to a disordered analogue of the $O(4)$ model with local gauge symmetry. Since gauge theory is applicable to the $SU(2)$ GG model, we examined the properties that hold on the NL using this expression, thereby validating our analysis.~\cite{Ozeki1993}\footnote{Due to errors in the reference, this expansion differs from the one given there.}

\subsection{Observables}
\label{subsec:observables}
    In this study, we simulate systems on a 3D lattice consisting of $N = L^2 \, (L-1)$ spins with an odd linear size $L$ under skew boundary conditions.
    Since the $O(4)$ model and the $SU(2)$ GG model are isotropic, it suffices to evaluate the physical observables using any single component of the four-component spin $\bm{S}_i =(\sigma_i, \bm{\pi}_i)$.

    Under these conditions, we perform Monte Carlo simulations in the vicinity of the expected transition point for several values of randomness $D$ to calculate the relaxation of magnetization $m(t)$ and that of SG susceptibility $\chi_\sg (t)$ as the targets of dynamical scaling analysis, providing the estimation of the FM or SG transition temperature.
    Then, we perform simulations at the FM transition point obtained above with a large number of samples to calculate the relaxation of FM susceptibility $\chi(t)$ and that of the covariance between magnetization and energy $\partial m(t)$, together with that of magnetization, providing precise estimations of FM critical exponents.
    We also perform simulations at the SG transition point obtained above, and calculate the SG correlation function $f_{\sg}(r,t)$ to estimate the relaxation of SG correlation length $\xi_\sg (t)$, providing the dynamical SG critical exponent.

    Starting from the initial state $\sigma_i = 1$, the magnetization $m$ is defined as
    \begin{equation}
        \label{eq:mag}
        \begin{aligned}
        \hat{m} &\equiv \frac{1}{N} \sum_{i=1}^N \sigma_i,\\
        m(t)&\equiv \configAve{\langle \hat{m}\rangle_t},
        \end{aligned}
    \end{equation}
    where $\langle\cdots\rangle_t$ denotes the dynamical average at time $t$, which is the average over dynamical samples.
    The susceptibility and the covariance between magnetization and energy are defined as
    \begin{equation}
        \label{eq:relaxation_chi}
    \begin{aligned}
          \chi(t) &\equiv \configAve{\langle \hat{m}^2\rangle_t - \langle \hat{m}\rangle_t^2},\\
        \partial m (t)&\equiv \Big|\configAve{\langle \hat{m}\hat{e}\rangle_t - \langle \hat{m}\rangle_t\langle \hat{e}\rangle_t}\Big|,
    \end{aligned}
    \end{equation}
    where $\hat{e}$ represents the energy per spin.
    The SG susceptibility $\chi_{\sg}$ is given by
    \begin{equation}
        \chi_{\sg}(t) = \frac{1}{N}\configAve{\sum_{ij} \langle \bm{S}_i \cdot \bm{S}_j \rangle_t^2 }.
        \label{eq:chisg}
    \end{equation}
    To calculate $\chi_{\sg}(t)$ practically in Monte Carlo simulations, we introduce $n$ replicas, which are copies of the system with the same disorder configuration.
    Letting $\bm{S}^{(A)}$ and $\bm{S}^{(B)}$ be the spins on replicas $A$ and $B$, we define the SG order parameter $q$ as
    \begin{equation}
        \label{eq:def_q}
        q_{\mu \nu}^{AB} \equiv \frac{1}{N} \sum_{i=1}^N \langle S_{i\mu}^{(A)} S_{i\nu}^{(B)}\rangle_t,
    \end{equation}
    where $\mu, \nu$ are the components of $\bm{S}_i$. The SG susceptibility is rewritten in terms of $q$ as
    \begin{equation}
        \label{eq:sus}
        \chi_{\sg} (t)= \frac{N}{\mathrm{C}_n} \ConfigAve{\sum_{A>B} \sum_{\mu, \nu} (q_{\mu \nu}^{AB})^2 },
    \end{equation}
    where $\mathrm{C}_n = n(n-1)/2$.
    Similarly, the SG correlation function defined by
    \begin{equation}
        \label{eq:sg_correlation}
            f_\sg(r,t) = \configAve{\frac{1}{N} \sum_{i=1}^{N} \langle \bm{S}_i \cdot \bm{S}_{i+r} \rangle_t^2 }
    \end{equation}
    is rewritten by the replica formula using the local SG order parameter $q_{\mu, \nu}^{AB}(i) \equiv S^{(A)}_{i \mu} S^{(B)}_{i \nu}$ as
    \begin{equation}
        \label{eq:correlation_function}
            f_{\sg}(r,t) = \frac{1}{N \mathrm{C}_n}\configAve{\sum_{A>B} \sum_{i,\mu, \nu}q^{AB}_{\mu\nu}(i)q^{AB}_{\mu\nu}(i+r)},
    \end{equation}
    where $r$ is the distance between spins on the lattice.
    However, in this case, it is numerically more practical to expand \cref{eq:correlation_function} as
    \begin{equation}
        \begin{aligned}
            f_{\sg}(r,t) = \frac{1}{2N \mathrm{C}_n}\Big[&\sum_{i=1}^N \Big\{ \Big(\frac{1}{n} \sum_{A=1}^n \bm{S}_i^{(A)} \cdot \bm{S}_{i+r}^{(A)}\Big)^2 \Big.\Big.
            \\
            &\Big.\Big. - \frac{1}{n^2} \sum_{A=1}^n \left(\bm{S}_i^{(A)} \cdot \bm{S}_{i+r}^{(A)}\right)^2 \Big\} \Big]_{\mathrm{c}}.
        \end{aligned}
    \end{equation}

\subsection{Results of gauge theory}
    The $SU(2)$ GG model \cref{eq:su2_hamiltonian} with \cref{eq:su2_probability_density} was first proposed in Ref.~\onlinecite{Ozeki1993}, where, based on gauge theory, several exact relations were derived, especially around the NL.
    In the present model, the NL is defined as
    \begin{equation}
    1/D=T.
    \label{eq:NL}
    \end{equation}
    Note that we recognize the temperature as a dimensionless parameter set, $k_{\mathrm{B}}T/J$, and use the variable $T$ in \cref{eq:NL} to represent this set hereafter.
    On this line, exact expressions for the internal energy, an upper bound of the specific heat providing no divergence, and exact identities for correlation functions can be derived regardless of the spatial dimension or lattice structure.
    It can be shown that the NL passes through the MCP but never enters the SG phase, and the phase boundary between the FM and SG phases below the MCP must be vertical or reentrant.

    Here, we show two remarkable relations, which will be used later for checking the validity of the calculations.
    The first is the exact expression of internal energy on the NL,
       \begin{equation}
        \label{eq:exact}
        E = -2N_\mathrm{B} J \; \frac{I_2(2D)}{I_1(2D)},
    \end{equation}
    where $N_\mathrm{B}$ is the number of bonds and $I_n(x)$ is the modified Bessel function.
    The second is the coincidence of the FM and SG correlation functions,
    \begin{equation}
    \label{eq:cm2cq2}
      \ConfigAve{\langle \bm{S}_i \cdot \bm{S}_j \rangle } = \ConfigAve{\langle \bm{S}_i \cdot \bm{S}_j \rangle^2 }
    \end{equation}
    which implies the coincidence of the FM and SG order parameters as $m=q$.
    Since this relation holds not only for static quantities~\cite{Ozeki1993} but also for dynamical quantities,~\cite{Yukiyasu_Ozeki_2003}
    the dynamical extension of \cref{eq:cm2cq2} provides
    \begin{equation}
    \label{eq:m2c2}
        M^2(t)=\chi_\sg (t),
    \end{equation}
    where
    \begin{equation}
    \label{eq:m2t}
    \begin{aligned}
        M^2(t) &= \frac{1}{N} \sum_{ij} \ConfigAve{ \langle \bm{S}_i \cdot \bm{S}_j \rangle_t }\\
               &= \frac{1}{N}\configAve{\Big\langle \Big(\sum_i \bm{S}_{i} \Big)^2 \Big\rangle_t },
    \end{aligned}
    \end{equation}
    is recognized as a dynamical FM order parameter, and $\chi_{\sg}(t)$ is defined in \cref{eq:chisg}.
    We will confirm this coincidence through a dynamical scaling analysis of both $M^2(t)$ and $\chi_\sg(t)$ on the NL.

\section{Numerical Method}
\label{sec:method}

We analyze the present model using the NER method,~\cite{ozeki2007nonequilibrium} which provides a powerful framework for analyzing equilibrium phase transitions through the non-equilibrium dynamical behavior of observables. Because it does not require the system to reach thermal equilibrium, it has been widely used in the analysis of SG and GG models.~\cite{PhysRevB.70.184417,PhysRevE.79.041138,PhysRevB.82.014427,doi:10.7566/JPSJ.92.074003}
    Depending on the target phase transition, an appropriate initial spin configuration must be chosen.
    For the FM-PM transition, we simulate the relaxation of the order parameter from an ordered state.
    Conversely, for the PM-SG transition, the relaxation starts from a disordered state.
    By applying the dynamical scaling law to these relaxation processes in the vicinity of the critical temperature $\Tc$ or $T_\sg$, we can precisely estimate the transition point.
    Subsequently, critical exponents are determined using techniques such as the analysis of the relaxation of fluctuations or correlation-length scaling, which will be discussed later.

\subsection{FM transition}
\label{subsec:FM}
    The magnetization $m(t)$ in the FM-PM transition asymptotically behaves as
    \begin{equation}
        \label{eq:relaxation_m}
        m(t) \sim \begin{cases}
            \exp(-t/\tau) & \quad T > \Tc, \\
            t^{-\beta/(z\nu)} & \quad T = \Tc, \\
            m_{\mathrm{eq}} & \quad T < \Tc,
        \end{cases}
    \end{equation}
    where $\tau$ is the relaxation time and $z$ is the dynamical critical exponent. The exponent $z$ is defined through the dynamical correlation length at $T=\Tc$ as $\xi(t) \sim t^{1/z}$.
    Using a scaling function $\Phi$, the dynamical scaling law for $m(t)$ is given by
    \begin{equation}
        \label{eq:mag_dynamical_scaling}
        m(t) \sim t^{-\beta/(z\nu)} \, \Phi\left(\frac{t}{|T - \Tc|^{-z\nu}}\right).
    \end{equation}
    By estimating appropriate values for $\Tc, \beta/(z\nu)$, and $z\nu$, the data for $T > \Tc$ and $T < \Tc$ collapse onto separate scaling curves.
    Note that a similar scaling can be applied to $M^2(t)$ in \cref{eq:m2t} with $2\beta/(z\nu)$ replaced by $\beta/(z\nu)$.

\subsubsection{Relaxation of fluctuations}
\label{subsubsec:analysis_fluctuation}
    In the analysis of the relaxation of fluctuations, we accurately simulate the asymptotic behavior at the critical temperature $\Tc$, which is predetermined by the dynamical scaling analysis.
    In addition to the decay of $m(t)$ described in \cref{eq:relaxation_m}, the susceptibility $\chi (t)$ and the covariance between magnetization and energy \cref{eq:relaxation_chi} are examined, whose asymptotic behaviors are expressed as
    \begin{equation}
        \label{eq:asympt_chi}
        \begin{aligned}
          \chi (t)&\sim t^{\gamma / (z\nu)}, \\
        \partial m (t)&\sim t^{(1-\beta)/(z\nu)}.
        \end{aligned}
    \end{equation}
    To evaluate the critical exponents obtained by the asymptotic behaviors of the above relaxation functions, we introduce the so-called local exponents for these quantities, defined as
    \begin{equation}
    \label{eq:llambda}
    \begin{aligned}
        \lambda_m(t)&\equiv -\frac{\partial \log m(t)}{\partial \log t},\\
        \lambda_\chi(t)&\equiv \frac{\partial \log \chi(t)}{\partial \log t},\\
        \lambda_{\partial m}(t)&\equiv \frac{\partial \log \,\partial m(t)}{\partial \log t}.
    \end{aligned}
    \end{equation}
    By combining these values and using the hyperscaling relation $d\nu =2\beta + \gamma$ ($d$ is the dimensionality of the lattice), we can derive the local exponents of each critical exponent as
    \begin{equation}
    \label{eq:lexp}
        \begin{aligned}
            z(t)&=\frac{d}{2\lambda_m(t) +\lambda_\chi(t)},\\
            \nu(t)&=\frac{1/z(t)}{\lambda_m(t) +\lambda_{\partial m}(t)},\\
            \beta(t)&=\frac{\lambda_m(t)}{\lambda_m(t) +\lambda_{\partial m}(t)},\\
            \gamma(t)&=\frac{\lambda_\chi(t)}{\lambda_m(t) +\lambda_{\partial m}(t)}.
        \end{aligned}
    \end{equation}
    Note that while the data in simulations consist of finite-time measurements, the system size is large enough that finite-size effects are negligible within the observation time window.
    Therefore, we obtain the asymptotic (precise) critical exponents by extrapolating the data to the limit $t \to \infty$.
    For this entire extrapolation procedure, we employ the comprehensive analysis framework utilizing GPR established in Ref.~\onlinecite{doi:10.7566/JPSJ.93.114001}.

\subsubsection{Asymptotic limit $t \to \infty$ of local exponents}
\label{subsubsec:gpr}
    For the critical exponents of the FM-PM transition, we need to extrapolate local exponents, $\beta(t)$, $\gamma(t)$, $\nu(t)$, and $z(t)$, in \cref{eq:lexp} to $t\to\infty$.
    In previous work,~\cite{doi:10.7566/JPSJ.93.114001} GPR and the Shanks transformation were introduced for this purpose.
    Gaussian process regression is a method that assumes data follow a multivariate normal distribution and predicts the underlying distribution by determining hyperparameters that maximize the likelihood function.
    Recently, techniques incorporating GPR into NER have been investigated.~\cite{PhysRevE.94.043312,doi:10.7566/JPSJ.93.114001}
    Since we also employ this method in the present study, we briefly review the framework.
    We consider datasets $\bm{X}$ and $\bm{Y}$ that are assumed to follow the scaling relation $\bm{Y} = \bm{\Phi}(\bm{X})$. The likelihood function $\mathcal{L}$ is then given by
    \begin{equation}
        \label{eq:likelihood}
        \mathcal{L} = \frac{1}{\sqrt{|2 \pi \Sigma|}} \exp \left(- \frac{1}{2} \transpose{(\bm{Y} - \bm{\Phi})} \Sigma^{-1} (\bm{Y} - \bm{\Phi}) \right),
    \end{equation}
    where $\Sigma$ denotes the covariance matrix. The optimal parameters are obtained by maximizing this likelihood function.
    The covariance matrix $\Sigma$ used in the analysis of the relaxation of fluctuations is given by
    \begin{equation}
        \label{eq:vcm_fluctuations}
        \begin{aligned}
            \Sigma_{ij}(\bm{\theta}) &= E_{i}^2 \delta_{ij} + \theta_0^2 + \theta_1^2 \exp \left( -\frac{(X_i - X_j)^2}{2\theta_2^2} \right) \\
                                     &= E_{i}^2 \delta_{ij} + K(X_i, X_j, \bm{\theta}),
        \end{aligned}
    \end{equation}
    where $E_i$ is the error of $Y_i$, $\bm{\theta}$ represents the hyperparameters, and $K$ denotes the kernel function, for which we employ the Gaussian kernel.
    We can predict $Y$ as
    \begin{equation}
       \label{eq:prediction}
        \ypredict(X) = \transpose{\bm{k}} \inverse{\Sigma(\thetaOpt)} \bm{Y},
    \end{equation}
    where \(\bm{k} = (k_{i})\),  \(k_{i}(X) = K(X_{i}, X, \thetaOpt)\), and $\bm{\thetaOpt}$ maximizes the likelihood defined in \cref{eq:likelihood}.
    We can also analytically estimate the derivative of $Y$ with respect to $X$ as
    \begin{equation}
      \label{eq:prediction-of-derivative}
       \frac{\partial \ypredict(X)}{\partial X} = \left(\frac{\partial \transpose{\bm{k}}}{\partial X}\right) \inverse{\Sigma(\thetaOpt)} \bm{Y}.
    \end{equation}
    Note that the simple derivative such as $\lambda_{y} = \partial \log(y) / \partial \log(t)$, ($y = \log(1/m)$, $\log \chi$, or $\log \partial m$) can be unstable at the boundaries of $t$.
    Instead of directly using the double-logarithmic derivative $\partial \log(y)/\partial \log(t)$, we convert the data into the following variables and perform the regression on $\partial Y / \partial X$:
    \begin{equation}
      \label{eq:converted-data}
      \begin{aligned}
        X_{i}
        &\equiv \frac{1}{\log(t_{i}) + c_{x}},\\
        Y_{i}
        &\equiv \frac{1}{c_{y1}\log(y_{i}) + c_{y2}},\\
        E_{i}
        &\equiv \frac{c_{y1}e_{y_{i}}}{y_{i}\left(c_{y1}\log(y_{i}) + c_{y2}\right)^{2}},\\
        c_{x}
        &\equiv 1 - \min_{i}\Set{\log(t_{i})},\\
        c_{y1}
        &\equiv \frac{\max_{i}\Set{\log(t_{i})} - \min_{i}\Set{\log(t_{i})}}{\max_{i}\Set{\log(y_{i})} - \min_{i}\Set{\log(y_{i})}},\\
        c_{y2}
        &\equiv 1 - c_{y1}\min_{i}\Set{\log(y_{i})},\\
      \end{aligned}
    \end{equation}
    where the index $i$ labels data points sequentially, $e_{i}$ represents the error in $y_{i}$, $E_i$ represents the error in $Y_i$, and $c_{x}$, $c_{y1}$, and $c_{y2}$ are constants determined by the data.
    This conversion has two advantages.
    First, $X_i$ and $Y_i$ are normalized to the range $0 \le X_{i}, Y_{i} \le 1$, corresponding to Min-Max Normalization.
    Second, at the critical temperature, the known limit $(X,Y) \to (0, 0)$ can be incorporated as a constraint.
    Therefore, we include the data point $(X_{i}, Y_{i}, E_{i}) = (0, 0, 0)$.
    The local exponents are then expressed as
    \begin{equation}
       \label{eq:prediction-of-lambda}
        \lambda_{y}(t) = \frac{X^{2}}{c_{y1}\left(\ypredict(X)\right)^{2}}\frac{\partial \ypredict(X)}{\partial X},
    \end{equation}
    where $\ypredict(X)$ and $\frac{\partial \ypredict(X)}{\partial X}$ are obtained from \cref{eq:prediction,eq:prediction-of-derivative}.
    We can interpolate the local exponents at $t$, such as $\beta(t)$, $\gamma(t)$, $\nu(t)$, and $z(t)$, by using $\lambda_{m}(t)$, $\lambda_{\chi}(t)$, and $\lambda_{\partial m}(t)$.

    The Shanks transformation~\cite{shanks1955non} is an extrapolation method that assumes the data points follow an exponential decay.
    In the computation, we use the $\varepsilon$-algorithm,~\cite{wynn1956device,sidi2003practical,brezinski2020extrapolation} which is an efficient implementation of the Shanks transformation.
    We assume that the local exponents converge exponentially and extrapolate them as $t \to \infty$ to obtain the critical exponents.

\subsection{SG transition}
\label{subsec:SG}
    In the case of the SG susceptibility $\chi_{\sg}$, the dynamical scaling law is given by
    \begin{equation}
        \label{eq:sus_dynamical_scaling}
        \chi_{\sg}(t) \sim t^{\gamma / (z_s\nu)} \, \Phi\left(\frac{t}{|T - T_{\sg}|^{-z_s\nu}}\right).
    \end{equation}
    It should be noted that for the PM-SG transition, the dynamical scaling law \cref{eq:sus_dynamical_scaling} holds for $T > T_{\sg}$. It has been pointed out that for the SG transition, $z_s$ takes the form:
    \begin{equation}
        z_s(T) = z \frac{T_{\sg}}{T}.
    \end{equation}
    This relation was shown to hold for 3D and 4D Gaussian-bond Ising SGs and the XY GG by Katzgraber \textit{et al}.~\cite{PhysRevB.72.014462}, and has since been confirmed in various SG systems.~\cite{nakamura2006nonequilibriumdynamicexponentspinglass,doi:10.7566/JPSJ.92.074003}
    By estimating appropriate values for $T_\sg, \gamma/(z\nu)$, and $z\nu$, the data for $T > T_\sg$ collapse onto their respective scaling curves.

    As for the two-point correlation function $f_\sg(r)$, it is known that the behavior at $T_\sg$ follows the power law:
    \begin{equation}
        f_\sg(r) \sim r^{-d + 2 - \eta}.
    \end{equation}
    Since the SG correlation length $\xi_\sg (t)$ and the distance $r$ between spins have the same scaling dimension, the following relation holds:~\cite{Nakamura_2016}
    \begin{equation}
        \label{eq:correlation_function_scaling}
        \frac{f_\sg(r,t)}{[\xi_\sg (t)]^{-d+2-\eta_{\mathrm{eff}}}} = \Phi \left(\frac{r}{\xi_\sg (t)} \right),
    \end{equation}
    where $\eta_{\mathrm{eff}}$ is the effective exponent. Thus, the values of $\xi_\sg (t)$ for every observed $t$ and those of $\eta_{\mathrm{eff}}$ are estimated such that the data collapse onto a single scaling curve.
    We performed the fitting procedure using a method recently developed via neural networks.~\cite{terasawa2026neural}

    Then, using the obtained values of $\xi_\sg (t)$ at $T=T_\sg$, which asymptotically behaves as
    \begin{equation}
        \xi_\sg (t)\sim t^{1/z},
    \end{equation}
    we estimate the dynamical exponent $z$ by introducing its local exponent as
    \begin{equation}
     \label{eq:def_z}
     z(t)\equiv \left\{\frac{\partial \log \xi_\sg (t)}{\partial \log t}\right\}^{-1}.
    \end{equation}
    The regression and extrapolation are performed in the same manner as analyzed for the FM transition in \cref{subsubsec:gpr}.

\subsection{Simulation procedure}
    In this study, we performed the analysis according to the following steps:

    \begin{itemize}
        \item[\textbf{a)}] \textbf{Sampling:}
        We computed the order parameters in the vicinity of the transition temperature ($\Tc$ or $T_\sg$) using Monte Carlo simulations. The system size $L$ was chosen to be large enough to ensure that finite-size effects are negligible.
        For the FM transition, we measured $m(t)$ up to $10^4$ Monte Carlo steps per spin (MCS), starting from an ordered state, as required by \cref{eq:mag}.

        In contrast, for the SG transition, the SG susceptibility $\chi_{\sg}$ was measured up to $2 \times 10^3$ MCS using 8 replicas, starting from a disordered state, based on \cref{eq:sus}.
        Additionally, each spin update in the SG transition consisted of one Metropolis step followed by one over-relaxation step.

        In both simulations, statistical averages were obtained from 1,024 independent samples.

        \item[\textbf{b)}] \textbf{Estimation of critical temperature:}
        We analyzed the obtained data based on the dynamical scaling laws given by \cref{eq:mag_dynamical_scaling} or \cref{eq:sus_dynamical_scaling}.
        We excluded the initial relaxation data from the analysis, as it deviates from the dynamical scaling laws.
        Statistical errors were evaluated using the bootstrap method with 128 resamplings.

        \item[\textbf{c)}] \textbf{Estimation of critical exponents:}
        For the FM transition, we performed simulations at $T=\Tc$ and calculated $m(t)$, $\chi(t)$, and $\partial m(t)$. As seen in \cref{eq:relaxation_m,eq:asympt_chi}, each observable asymptotically decays algebraically with $t$ to the power of the critical exponents. We can obtain each exponent by combining the asymptotic behaviors of these observables.
        Statistical averages were taken over approximately $10^6$ samples.

        Using the scale in \cref{eq:converted-data} and performing GPR, we evaluated the local exponent in \cref{eq:llambda} for each observable from the derivative defined in \cref{eq:prediction-of-derivative}, providing the estimation of local exponents $\beta(t)$, $\gamma(t)$, $\nu(t)$, and $z(t)$ defined in \cref{eq:lexp}.
        To apply the $\varepsilon$-algorithm~\cite{wynn1956device,sidi2003practical,brezinski2020extrapolation} to the data obtained above, we divided the data into $2k+1$ equally spaced points and applied the $\varepsilon$-algorithm $k$ times to these data points.
        The median of the final terms of the resulting $k$ sequences was then adopted as the estimate of the critical exponent.
        In the present study, we used $k = 500$.

        As for the SG transition, we calculated the SG correlation function $f_{\sg}(r,t)$ at $T_{\sg}$, then performed the correlation-length scaling and obtained a data collapse according to \cref{eq:correlation_function_scaling}.
        Since the dynamic critical exponent $z$ can be evaluated from $\xi_\sg (t)$ via \cref{eq:def_z}, we determined the exponent $\nu$ by combining it with the scaling parameters $z\nu$ obtained from \cref{eq:sus_dynamical_scaling}.
        Furthermore, we estimated the exponent $\eta$ using the scaling relation $\eta = 2 - \gamma/\nu$, where the ratio $\gamma/\nu$ was also derived from \cref{eq:sus_dynamical_scaling}.
        Errors were evaluated using the bootstrap method with 100 resamplings.
        It is noted that the evaluation of critical exponents using scaling parameters may involve systematic errors due to the finite observation time. Nevertheless, we choose to use them in the case of SG because, in this case, the computational cost of measuring the relaxation of fluctuations is too high.
    \end{itemize}

\section{Results}
    \label{sec:result}

    \subsection{Finite-size effects}
\begin{figure}[htbp]
        \centering
        \includegraphics[width=6cm]{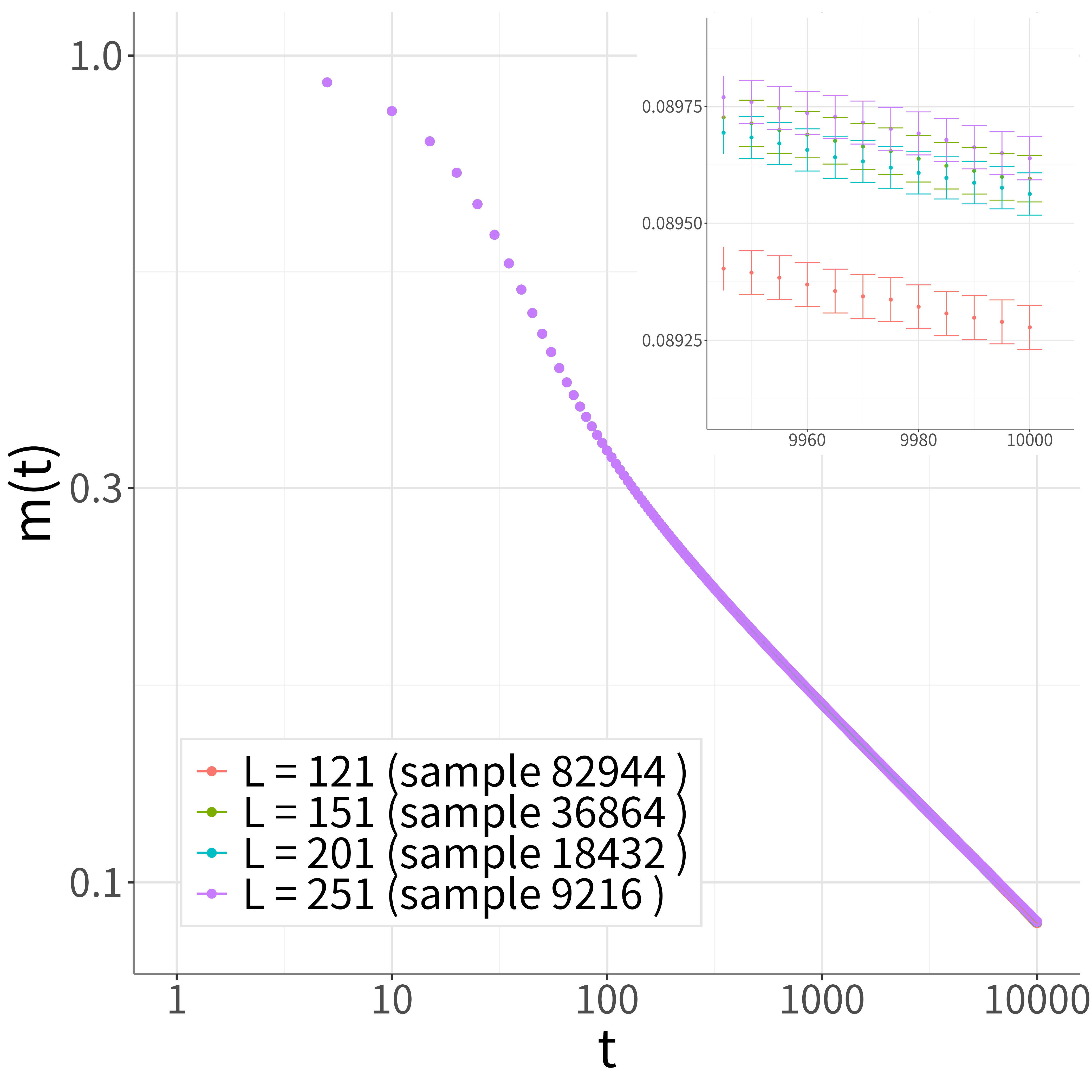}
    \caption{\label{figure:mag_finite_size_effect}
     Size dependence of $m(t)$ as a function of $t$ on a double-logarithmic scale at $T=0.728$ for $1/D=0$.
     The inset shows an enlarged view in the range $9950 \le t \le 10000$.
     For $L \ge 151$, the data overlap within statistical errors, indicating that the size dependence is negligible.}
    \end{figure}

\begin{figure}[htbp]
        \centering
        \includegraphics[width=6cm]{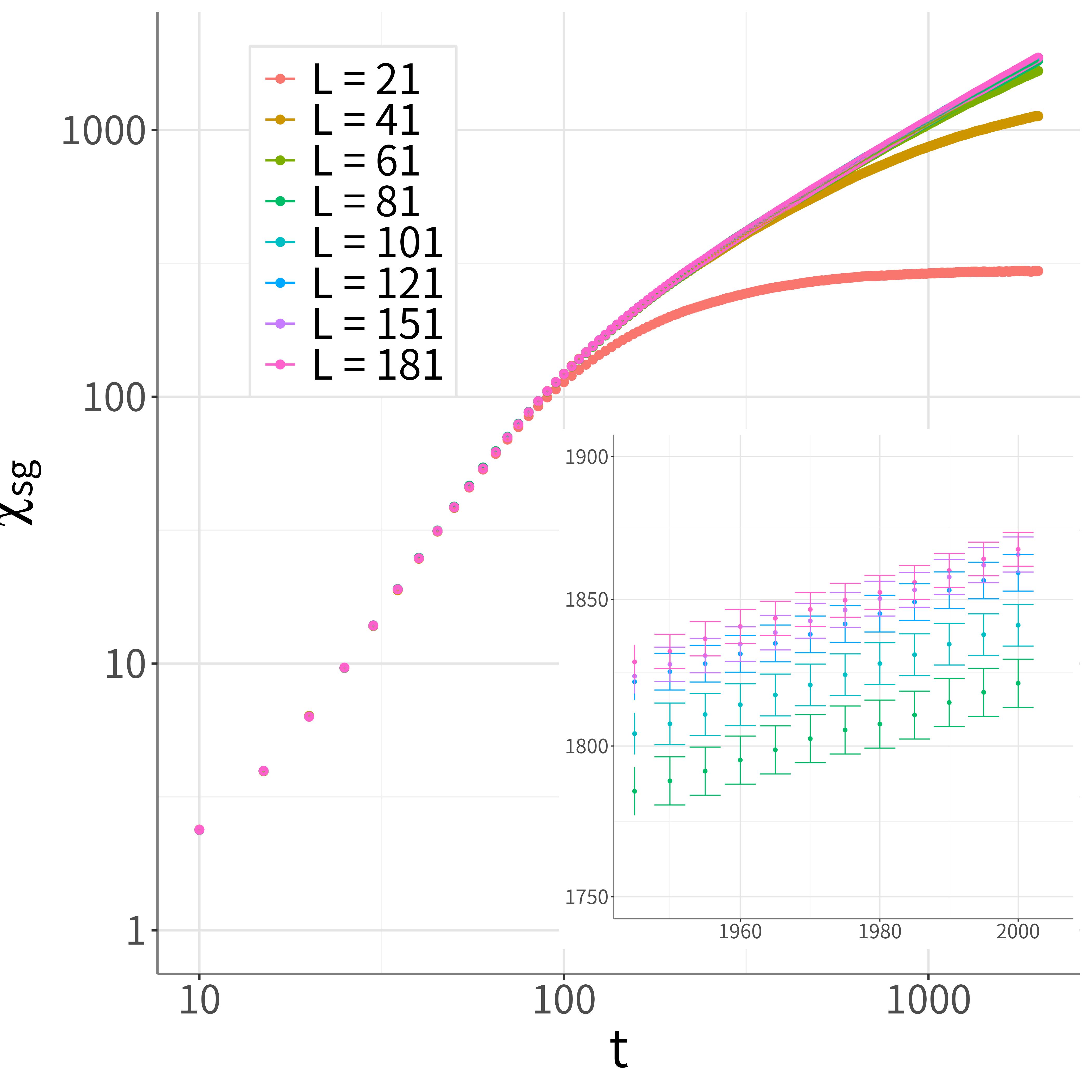}
        \caption{\label{figure:sus_finite_size_effect}
            Size dependence of $\chi_{\sg}(t)$ as a function of $t$ on a double-logarithmic scale at $T = 0.46$ for $1/D = \infty$, which is close to the SG transition point.
            The inset shows an enlarged view of the data in the range $1950 \le t \le 2000$.
            For $L \ge 121$, the data overlap within statistical errors, indicating that the size dependence is negligible. }
    \end{figure}

    In NER analysis, it is necessary to evaluate finite-size effects because the method is intended to describe the behavior of infinite systems. We first examined the finite-size effects on the magnetization $m(t)$, the SG susceptibility $\chi_{\sg}$, and the SG correlation function $f_{\sg}(r,t)$.
    \Cref{figure:mag_finite_size_effect} shows the size dependence of $m(t)$ at an expected FM transition point, and \cref{figure:sus_finite_size_effect} shows the size dependence of $\chi_{\sg}(t)$ at an expected SG transition point.
    To avoid deviations in the relaxation functions due to finite-size effects, we employed systems large enough that the size dependence becomes negligible.


\begin{figure}[htbp]
        \centering
        \includegraphics[width=6cm]{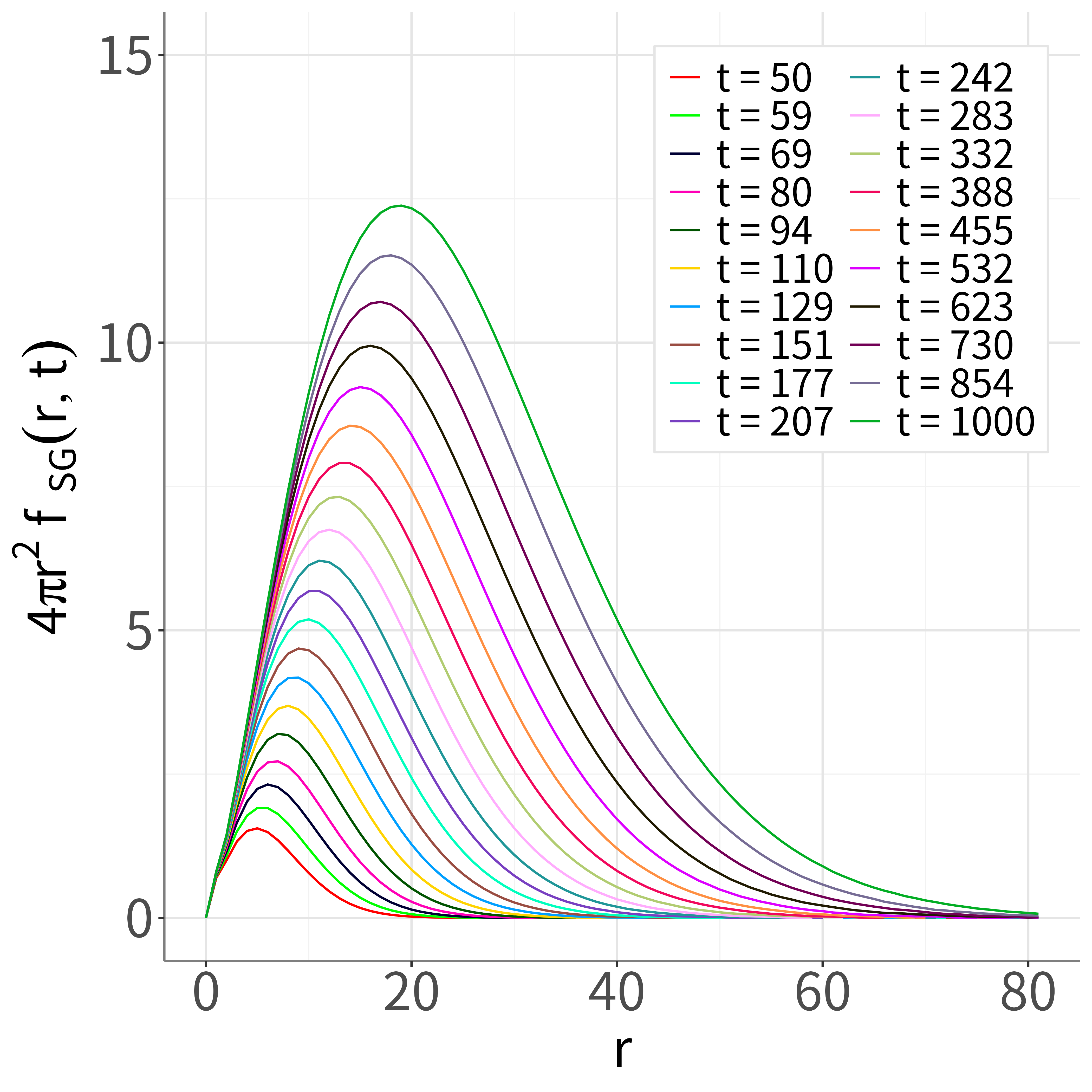}
        \caption{\label{figure:corr_finite_size_effect}
            Weighted SG correlation function $4 \pi r^2 f_{\sg}(r, t)$ at the SG critical temperature ($T_\sg = 0.4504$) for $1/D = 2.0$.
            Data for $L = 251$ are shown within the range $0 \le r \le L/3$.
            The asymptotic convergence to zero indicates that finite-size effects, which would otherwise appear as upward deviations, are negligible.}
    \end{figure}

    \Cref{figure:corr_finite_size_effect} shows the SG correlation function $f_{\sg}(r,t)$ for $L=251$.
    A previous study~\cite{Nakamura_2019} pointed out that $4\pi r^2 f_{\sg}(r,t)$ deviates upward for small $L$ and converges to zero for sufficiently large systems.
    We confirmed that such a finite-size effect is absent in the system sizes adopted in our simulations.
    These procedures were similarly applied to the models with other degrees of disorder to perform simulations at appropriate sizes.

\subsection{Checking validity}
   In this subsection, we verify the validity of our analysis through two comparisons on the NL, which are based on \cref{eq:exact,eq:cm2cq2}.
    As the first verification, we consider the internal energy on the NL, \cref{eq:exact}.
    Therefore, we calculated the thermal average of it by performing Monte Carlo simulations with $1000$ MCS averaging for $L=41$, using 1024 independent samples.
    From the time evolution of the magnetization, we confirmed that the system undergoes sufficient spin flips and that the internal energy remains stable near the thermal average.
    In \cref{table:energy}, numerical results for the internal energy are compared with the exact values from \cref{eq:exact}.
    As shown in \cref{table:energy}, the numerical results are in excellent agreement with the exact values, confirming the reliability of our simulation code.

    \begin{table}[htbp]
        \centering
        \caption{Comparison of internal energies on the NL ($1/D=T$).
        The energy is measured in units of $J$.
        Exact values are calculated by \cref{eq:exact}, and numerical ones are estimated by Monte Carlo simulations.}
        \label{table:energy}
        \vspace{2mm}
        \centering
        \begin{tabular}{cll}
        \hline
        \hline
          $T$ & \cref{eq:exact} & numerical \\
        \hline
            0.1 & $-5.555927$  & $-5.555923(5)$ \\
            0.5 & $-3.94828$ & $-3.94827(2)$ \\
            1.0 & $-2.59876$ & $-2.59876(5)$ \\
            2.0 & $-1.44116$ & $-1.44116(5)$ \\
        \hline
        \hline
        \end{tabular}
    \end{table}

    Next, as the second verification, we examine the identity \cref{eq:m2c2} provided by \cref{eq:cm2cq2} on the NL.
    They ensure that the FM and SG correlations, as well as their order parameters, are identical on the NL, which reveals that the MCP can be estimated both by the dynamical scaling of $M^2(t)$ and that of $\chi_\sg(t)$.
    We performed simulations starting from a completely disordered state using sufficiently large lattice sizes to neglect finite-size effects.
    For $M^2(t)$, the data were averaged over 27,648 samples, while for $\chi_{\sg}(t)$, the averages were taken over 2,048 samples with 8 replicas each. In all simulations, we used $L = 151$ up to $t = 2000$ MCS.

\begin{figure}[htpb]
        \centering
            \centering
            \includegraphics[width=6cm]{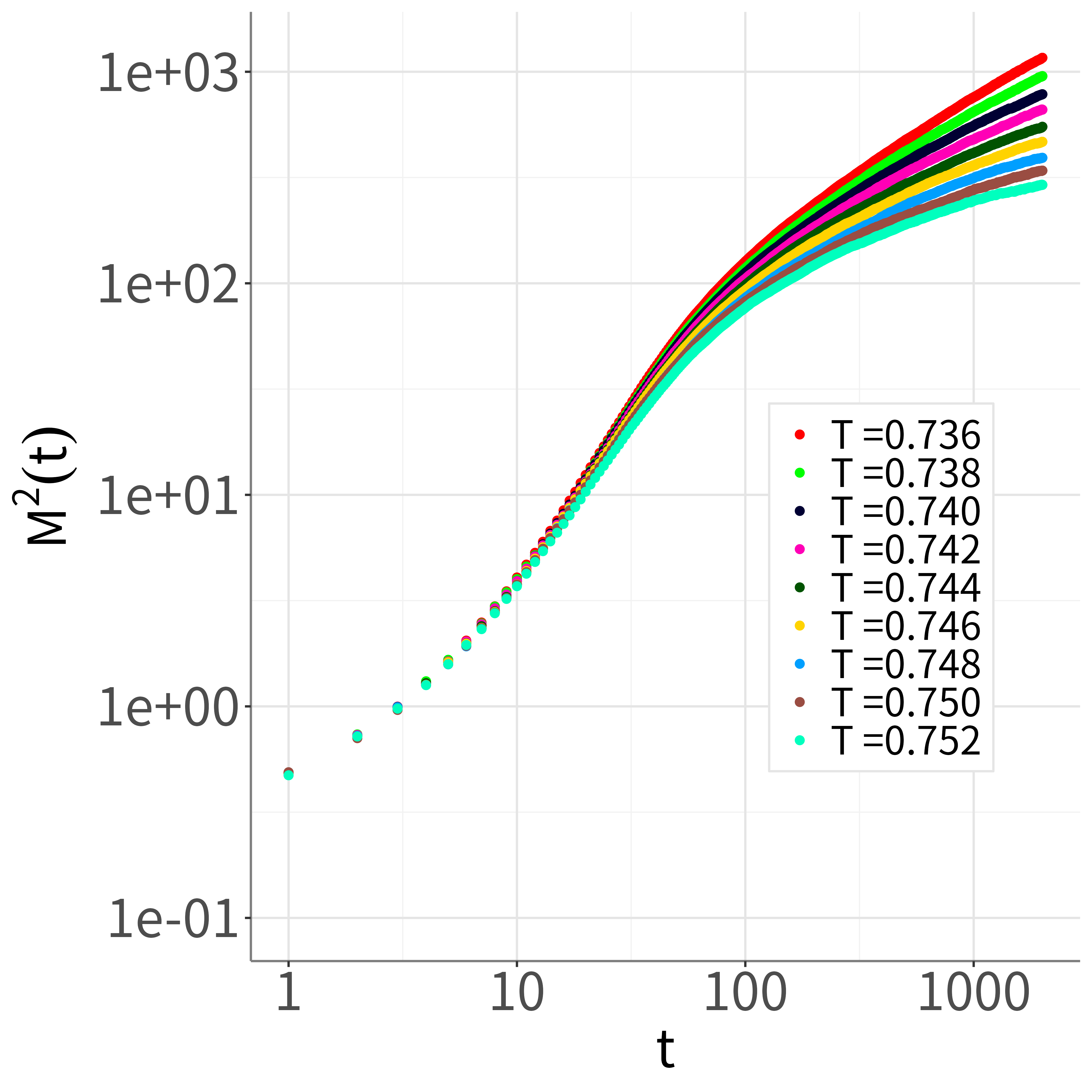}
            \centerline{(a)}
            \centering
            \includegraphics[width=6cm]{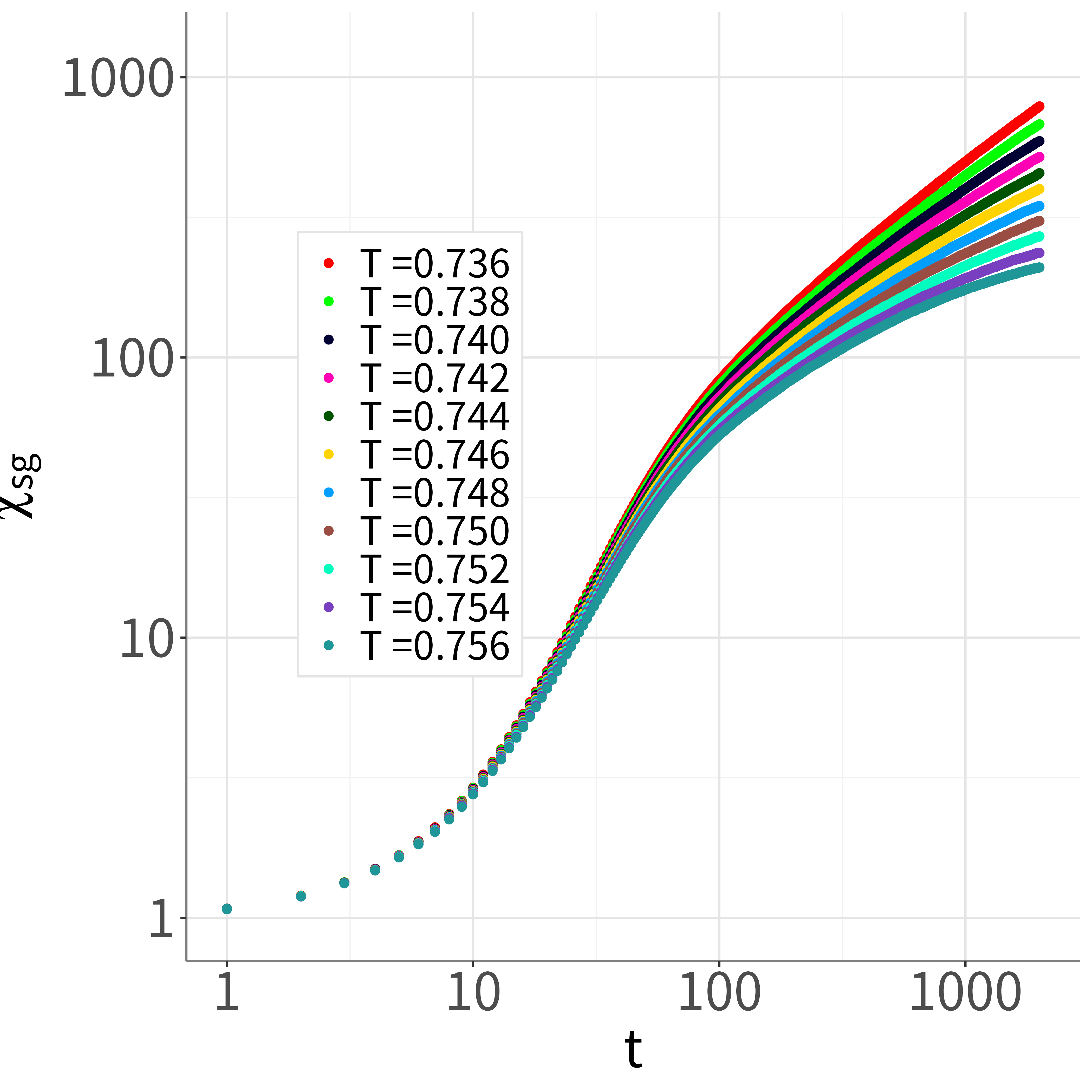}
            \centerline{(b)}
        \caption{\label{figure:scaling_mq}
        (a) Relaxation of $M^2(t)$ and (b) that of $\chi_\sg (t)$
        on the NL on a double-logarithmic scale, both calculated for $L = 151$ up to $t = 2000$ MCS. The dynamical scaling of \cref{eq:sus_dynamical_scaling} provides the transition temperature $T_{\fm} = 0.7284(2)$ from $M^2(t)$ and $T_\sg = 0.72865(8)$ from $\chi_\sg(t)$.
        }
    \end{figure}

    \Cref{figure:scaling_mq} shows the results for $M^2(t)$ and $\chi_\sg (t)$ above the transition temperatures.
    They almost coincide, although small numerical discrepancies are observed between them.
    These discrepancies are due to the absence of the diagonal part of the two replicas in \cref{eq:sus}, which should be negligible for sufficiently large $n$ and does not affect the subsequent scaling analysis.
    We made corresponding scaling plots for \cref{eq:sus_dynamical_scaling} and obtained the estimation of critical temperatures as $T_{\fm} = 0.7284(2)$ and $T_{\sg} = 0.72865(8)$; for the $M^2(t)$ case, the exponents and transition temperature are considered to be those for the FM transitions.
    These values are consistent with each other within statistical errors.
Furthermore, in the next subsection, we will estimate the FM transition temperatures by using the dynamical scaling law \cref{eq:mag_dynamical_scaling} for the magnetization $m(t)$, and the estimated $\Tc$ at the MCP is consistent with the above set of transition temperatures.

    \subsection{FM-PM boundary}
    \label{subsec:pf_transition}
    As mentioned in the introduction, for the $O(4)$ model, the specific heat exponent $\alpha$ is negative. According to the Harris criterion, weak disorder is expected to be irrelevant to the universality class for this system. In contrast, at the MCP where the Harris criterion is not applicable, the emergence of a different universality class is expected.

\begin{figure}[htbp]
        \centering
        \includegraphics[width=6cm]{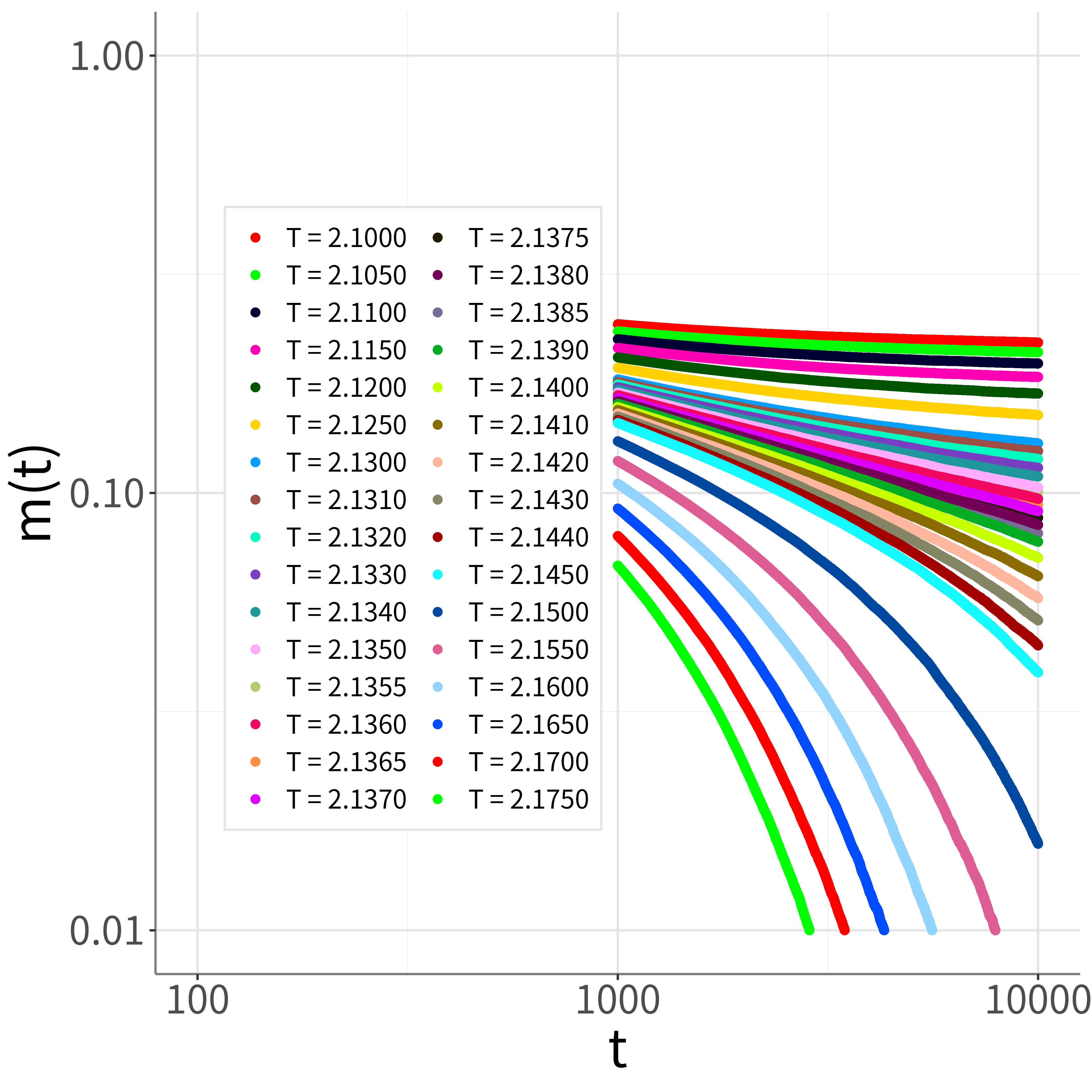}
        \centerline{(a)}
        \centering
        \includegraphics[width=6cm]{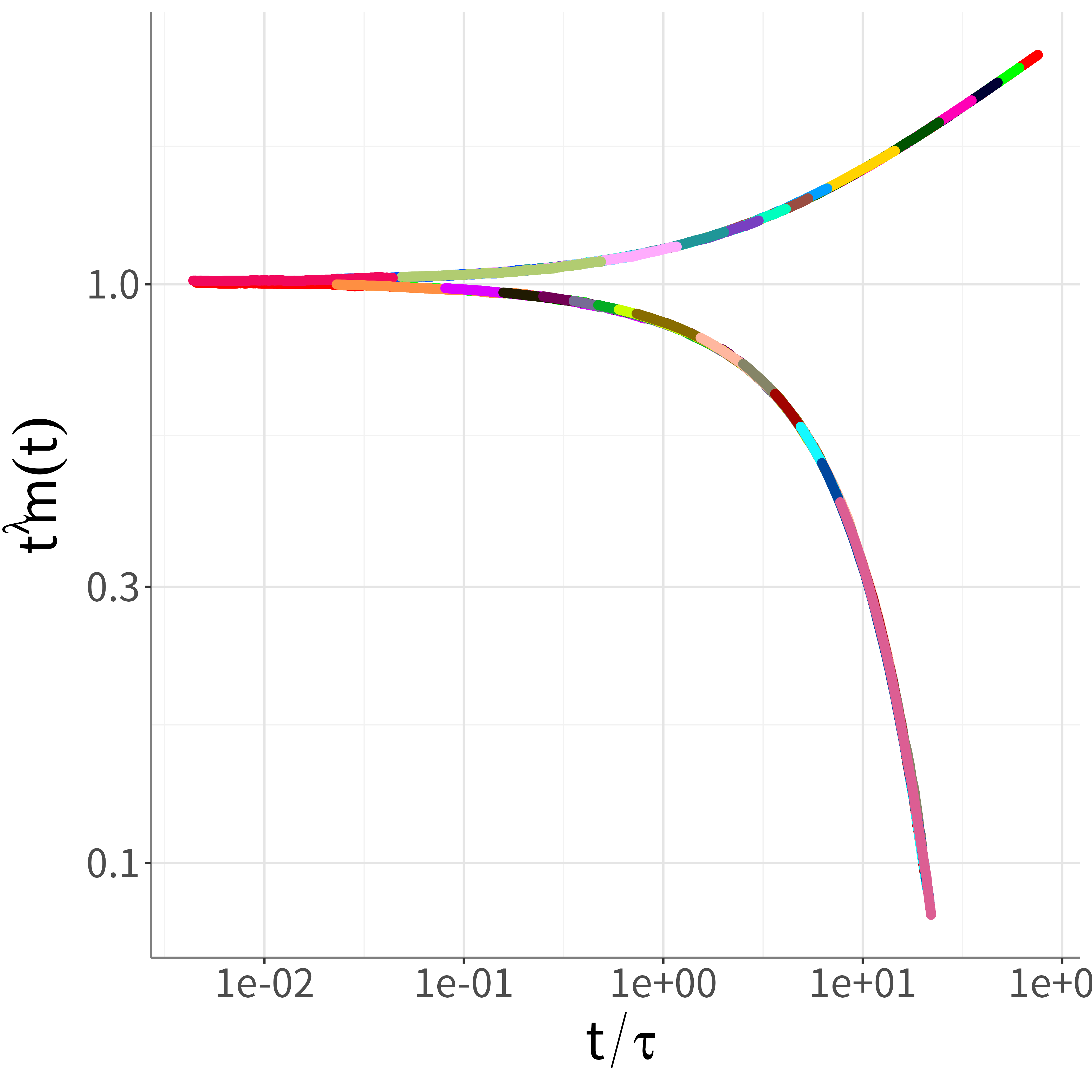}
        \centerline{(b)}
        \caption{\label{figure:mag_relaxation}
            (a) Relaxation of $m(t)$ for the $O(4)$ model ($1/D=0$) on a double-logarithmic scale for $L=151$ in the vicinity of $\Tc$, covering the range $2.100 \le T \le 2.175$.
            (b) Corresponding scaling plot according to \cref{eq:mag_dynamical_scaling}.
            The initial relaxation data for $t < 1000$ are excluded from the scaling analysis.
            The estimated parameters are $\Tc = 2.137247(3)$, $\beta / (z\nu) = 0.2632(9)$, and $z\nu = 1.484(2)$.}
    \end{figure}

    To demonstrate our scaling analysis procedure, we first present the detailed results for the pure system ($1/D=0$) as a representative case.
    \Cref{figure:mag_relaxation}(a) shows the relaxation data for $T=2.100$ to $2.175$ with $L=151$. Performing dynamical scaling plot according to \cref{eq:mag_dynamical_scaling}, we obtained the data collapse shown in \cref{figure:mag_relaxation}(b), where we obtained fitting parameters as $\Tc=2.137247(3)$, $\beta/(z\nu)=0.2632(9)$, and $z\nu=1.484(2)$.

\begin{figure}[htbp]
        \centering
            \centering
            \includegraphics[width=5.5cm]{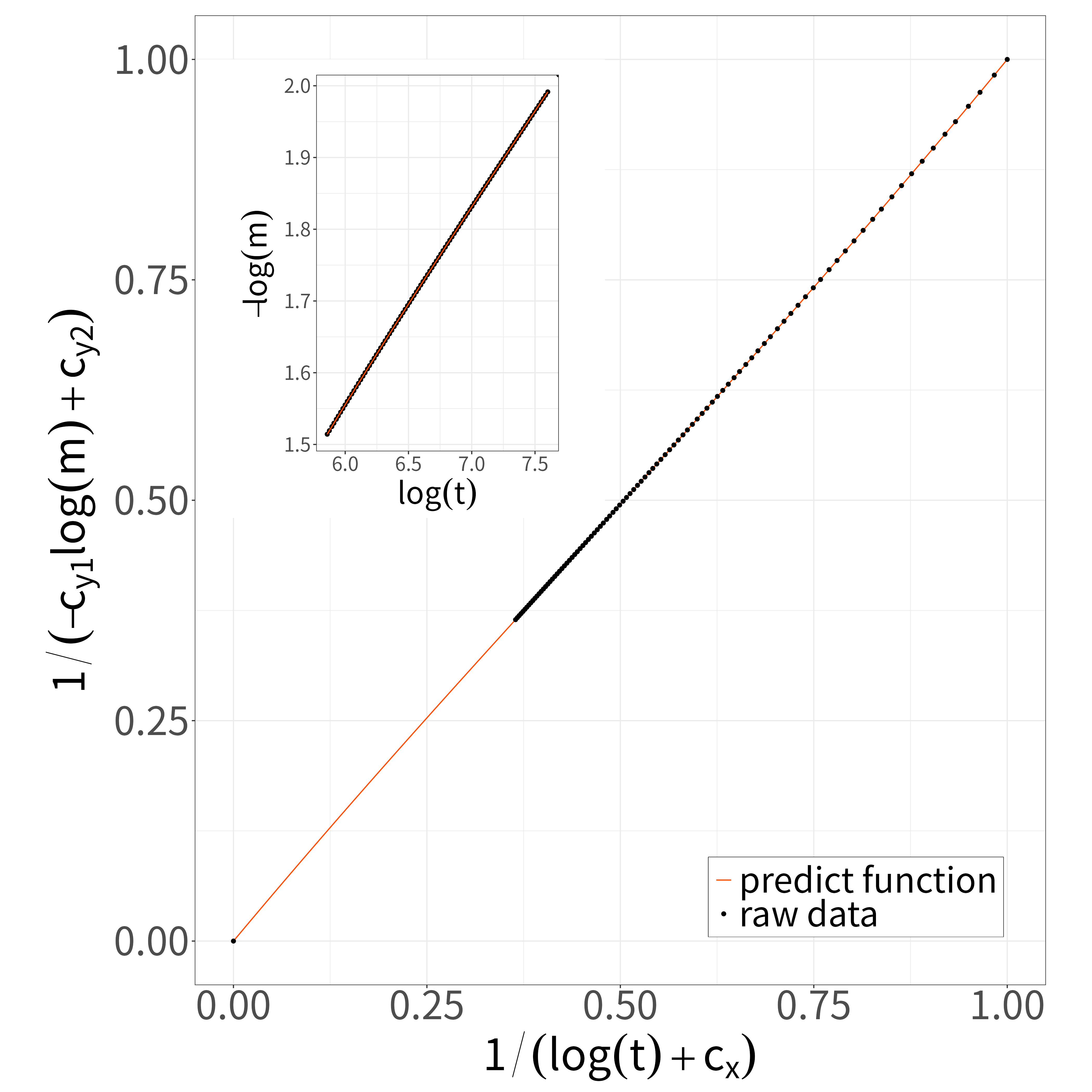}
            \centerline{(a)}
            \centering
            \includegraphics[width=5.5cm]{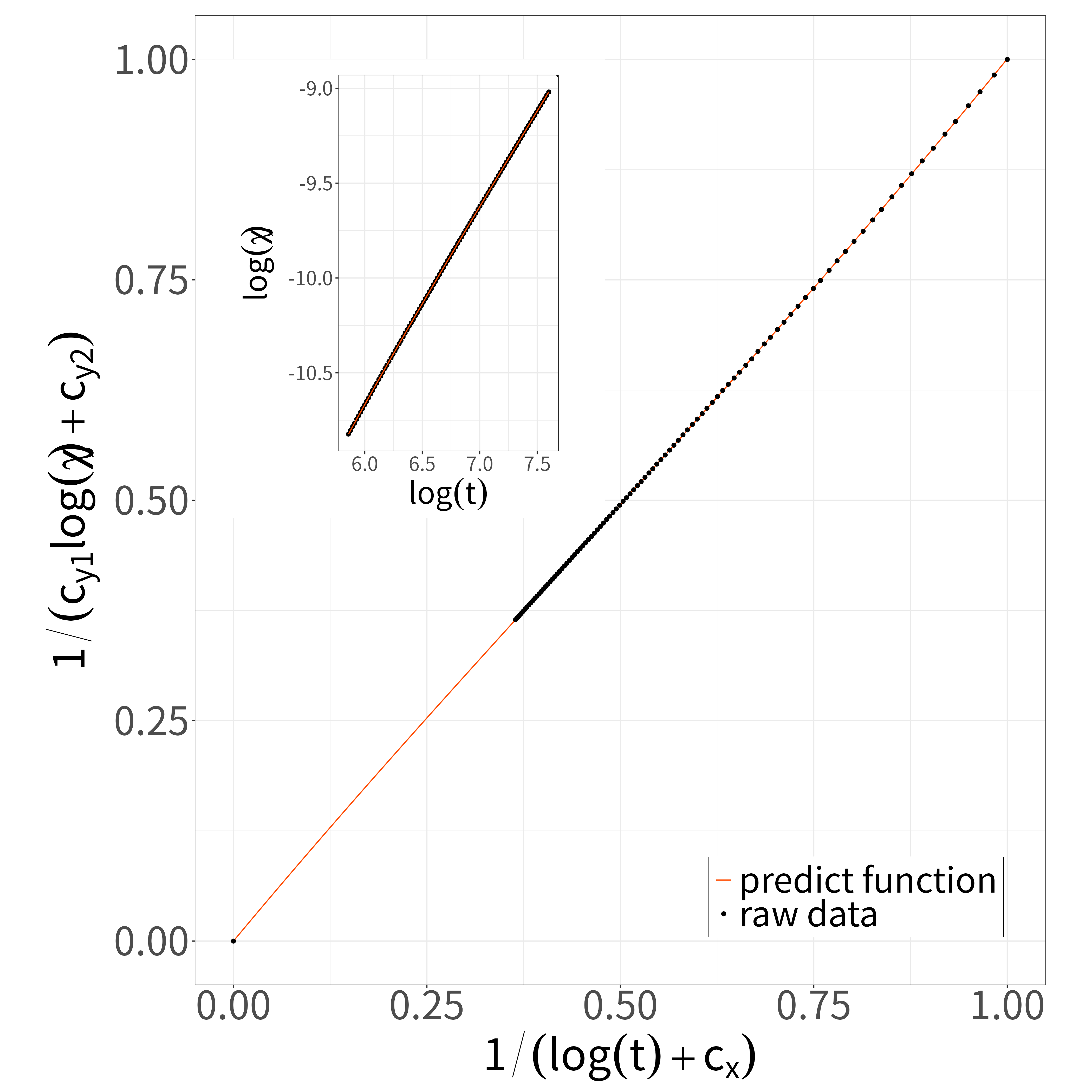}
            \centerline{(b)}
            \centering
            \includegraphics[width=5.5cm]{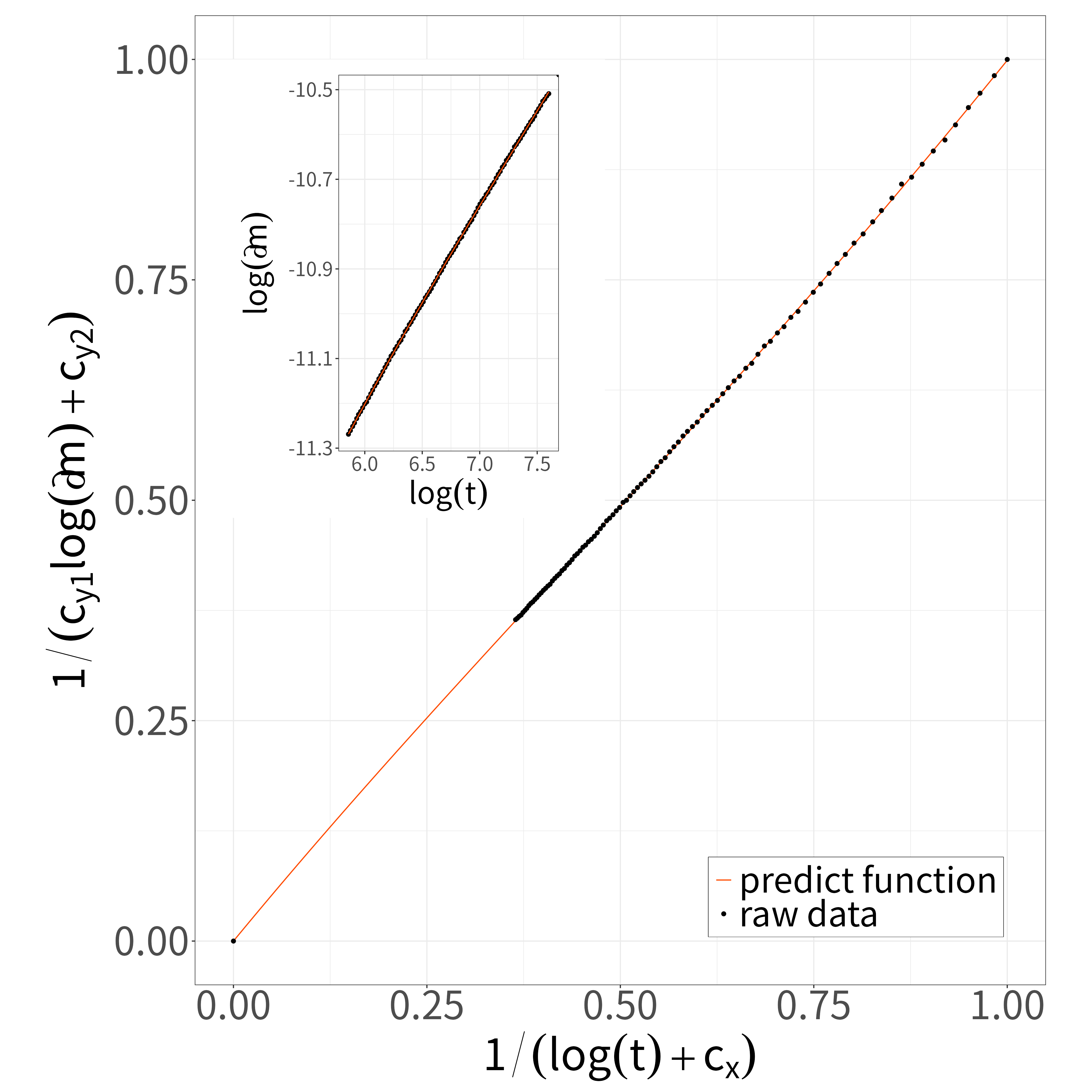}
            \centerline{(c)}
        \caption{\label{figure:regression_local}
        Local exponents (a) $\lambda_m(t)$, (b) $\lambda_\chi(t)$, and (c) $\lambda_{\partial m}(t)$ at $T=\Tc$ for the $O(4)$ model ($1/D=0$) plotted in the scale defined in \cref{eq:converted-data}.
        The horizontal axis represents the inverse of the logarithm of $t$, and the vertical axis represents the inverse of the logarithm of each physical quantity.
        The orange curves denote the regressions performed using data sampled at equal logarithmic intervals, along with the data point (0,0) corresponding to the thermodynamic limit.
        The simulation data (black dots) align well with the regression curves, demonstrating that the fitted curves accurately reproduce the raw data.}
    \end{figure}

\begin{figure}[htbp]
        \centering
           \centering
           \includegraphics[width=5cm]{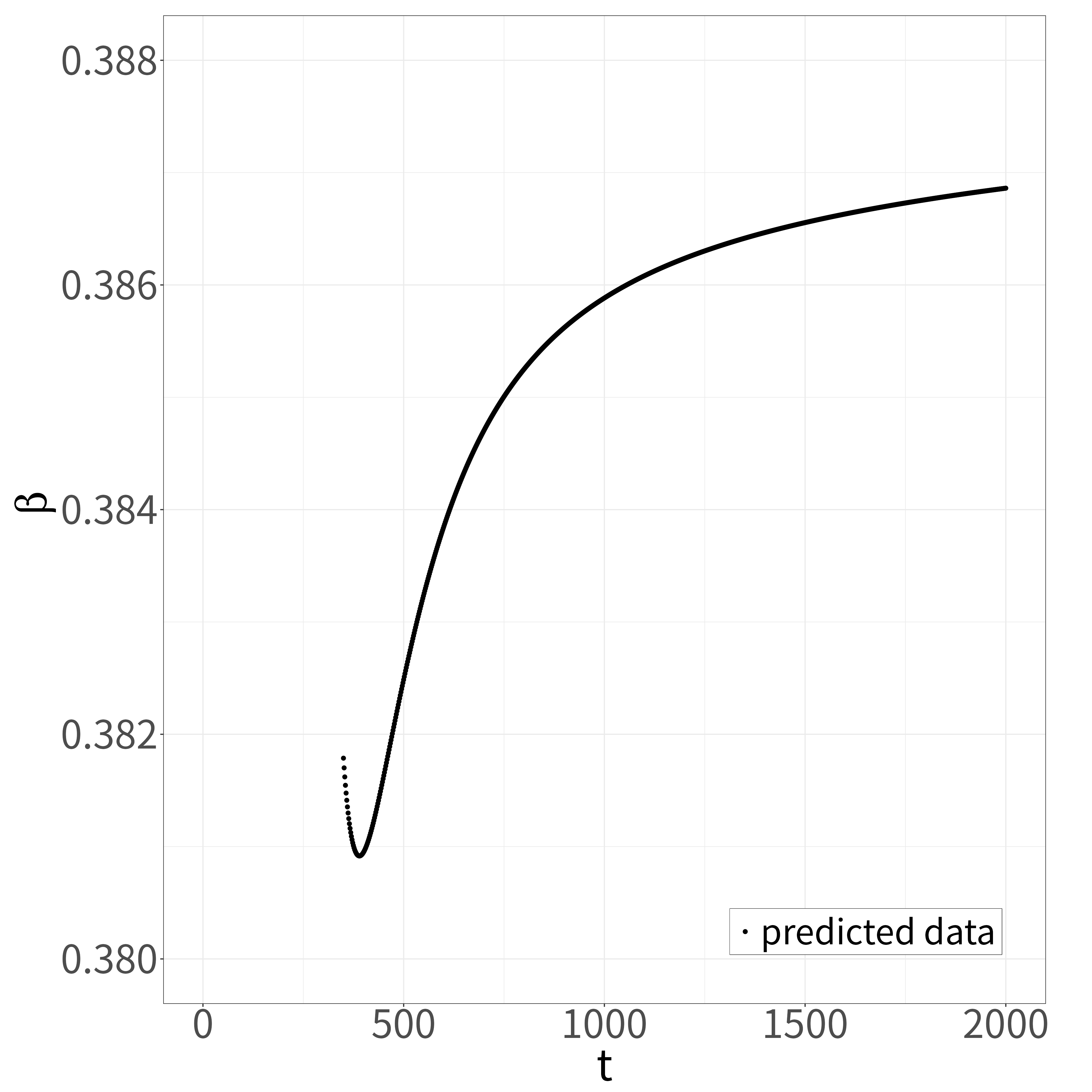}
           \centerline{(a)}
           \centering
           \includegraphics[width=5cm]{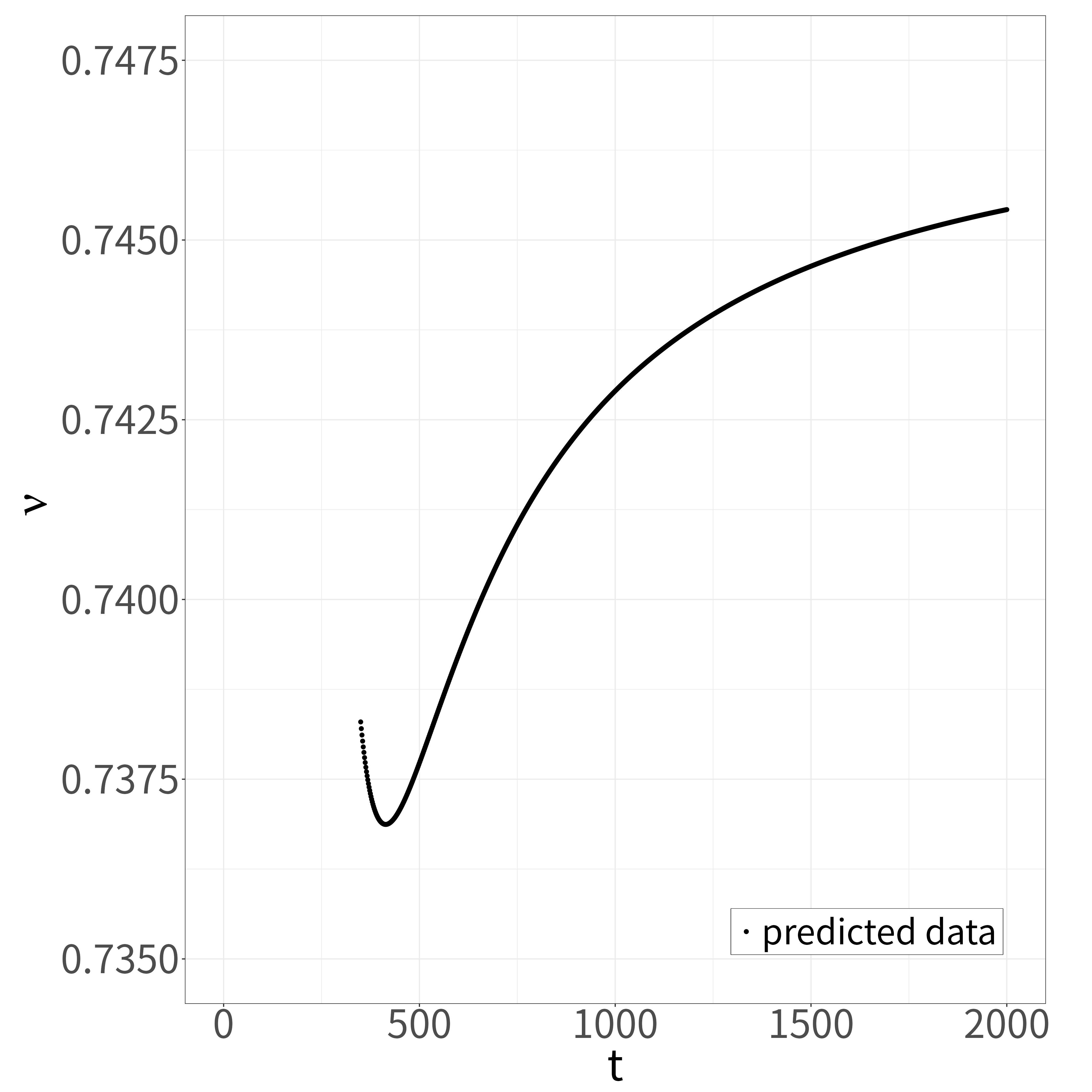}
           \centerline{(b)}
           \centering
           \includegraphics[width=5cm]{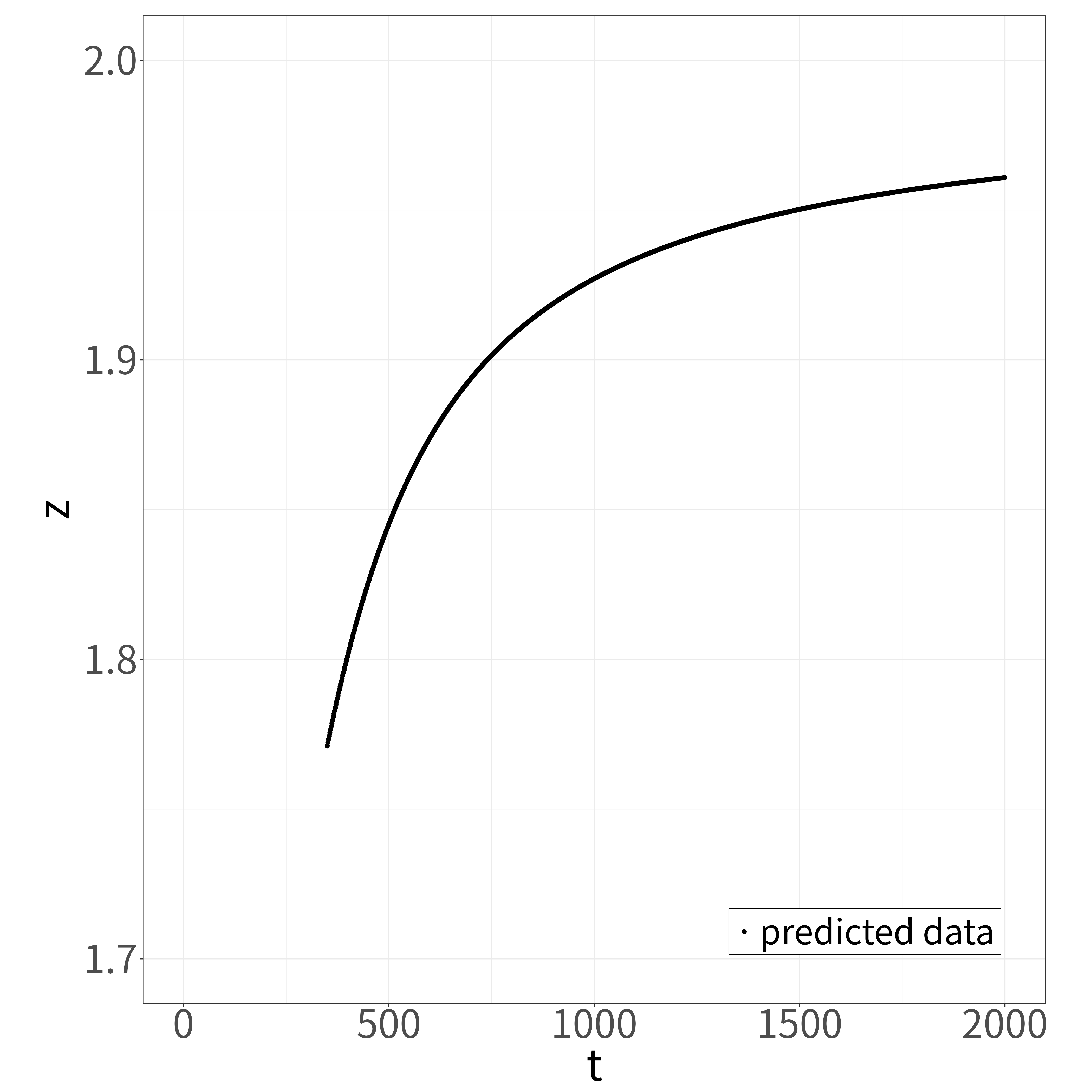}
           \centerline{(c)}
        \caption{\label{figure:regression_exponent}
            Local exponents (a) $\beta(t)$, (b) $\nu(t)$, and (c) $z(t)$ defined by \cref{eq:llambda,eq:lexp} calculated from the derivatives of the regression curves in \cref{figure:regression_local}. The regression range is set to $t \ge 350$.}
    \end{figure}

    Subsequently, we performed simulations in order to calculate $m(t)$, $\chi(t)$, and $\partial m(t)$ at $T=\Tc$ estimated above with $L=81$ up to $2000$ MCS for approximately $10^6$ samples.
    The results are plotted in \cref{figure:regression_local} on the converted scale of \cref{eq:converted-data}, where regression curves obtained by \cref{eq:prediction} are also shown.
    The local exponents in \cref{eq:llambda} are evaluated by these plots using the derivative \cref{eq:prediction-of-derivative}
    providing the estimation of local exponents $\beta(t)$, $\gamma(t)$, $\nu(t)$, and $z(t)$ defined in \cref{eq:lexp}.
    The results are shown in  \cref{figure:regression_exponent}.
    Extrapolating the obtained results to $t \to \infty$ yields the critical exponents $\nu = 0.7465(14)$, $\beta = 0.3868(8)$, $\gamma = 1.465(3)$, and $z = 1.982(1)$.
   Furthermore, using the Fisher relation $\eta = 2 - \gamma / \nu$, we evaluate $\eta = 0.0363(9)$.

    To confirm the validity of our analysis, in \cref{table:o4_comparison}, these estimates are compared with previous studies for the $O(4)$ model.
    The values obtained in this work are highly consistent with previous studies, establishing the reliability of our analysis.

    \begin{table}[htpb]
        \caption{Results for $\Tc$ and critical exponents for the 3D $O(4)$ model. They are almost consistent with previous studies; the methods are indicated as Monte Carlo (MC), conformal bootstrap (CB), renormalization group (RG), and non-equilibrium relaxation (NER).
        Values marked with an asterisk (*) are multiplied by 2 to account for the difference in the energy scale relative to the current work.}
        \label{table:o4_comparison}
        \vspace{2mm}
        \centering
        \begin{tabular}{cc|lll}
        \hline
        \hline
            & & $\Tc$ & $\nu$ & $\eta$ \\
        \hline
        Ref.~\onlinecite{PhysRevD.51.2404} & MC & \(2.1367(26)\)* & \(0.7479(90)\) & \(\) \\
            Ref.~\onlinecite{Kos2015} & CB & \(\) & \(0.7472(87)\) & \(0.0378(32)\) \\
           Ref.~\onlinecite{PhysRevE.104.064101} & fRG & \(\) & \(0.7478(9)\) & \(0.0360(12)\) \\
            Ref.~\onlinecite{PhysRevB.105.054428} & MC & \(\) & \(0.74817(20)\) & \(0.03624(8)\) \\ \hline
            This work & NER &  \(2.137247(3)\) & \(0.7465(14)\) & \(0.0363(9)\) \\
        \hline
        \hline
        \end{tabular}
    \end{table}

    Following the same procedure, we evaluated $\Tc$ and the critical exponents for various values of $1/D$ up to the NL ($1/D=T$), as summarized in \cref{table:critical_exponent}.
    In \cref{table:critical_exponent}, while $\Tc$ decreases monotonically with increasing $1/D$, the static critical exponents are consistent among the cases of $1/D=0$, $0.2$, and $0.5$.
    This strongly suggests that these cases belong to the same universality class.
    On the other hand, $z$ increases monotonically with $1/D$, confirming that the dynamics depends on the degrees of disorder.
    In contrast, the values at the MCP on the NL ($1/D=T$) deviate from those at other points, suggesting that this point belongs to a different universality class.

    \begin{table}[htpb]
    \centering
    \caption{Results for FM critical exponents.
    For $1/D=0$, $0.2$, and $0.5$, the static critical exponents agree within statistical errors, indicating that these systems belong to the same universality class.
    The bottom row ($1/D=T$) shows the results at the MCP on the NL, suggesting that it belongs to a different universality class.}
    \label{table:critical_exponent}
    \vspace{2mm}
    \centering
    \begin{tabular}{c|llll}
    \hline
    \hline
    $1/D$ & $\Tc$ & $\nu$ & $\beta$ & $z$ \\
    \hline
        0 & \(2.137247(3)\) &  \(0.7465(14)\) &  \(0.3868(8)\) & \(1.982(1)\) \\
        0.2 & \(1.78676(7)\) & \(0.7476(20)\) & \(0.3858(10)\) & \(2.041(1)\) \\
        0.5 & \(1.26776(4)\) & \(0.7482(9)\) & \(0.3875(7)\) & \(2.112(1)\) \\ \hline
        $T$ & \(0.7284(2)\) & \(0.987(1)\) & \(0.4235(7)\) & \(2.767(1)\) \\
    \hline
    \hline
    \end{tabular}
    \end{table}

    \subsection{PM-SG boundary}
    \label{subsec:pm_sg_transition}
    We demonstrate the existence of the SG phase at finite temperatures in the $SU(2)$ GG model, for which no previous studies exist.
    We also verify the hypothesis that the SG transition possesses a universality independent of the disorder level.

\begin{figure}[htbp]
            \centering
                \includegraphics[width=6cm]{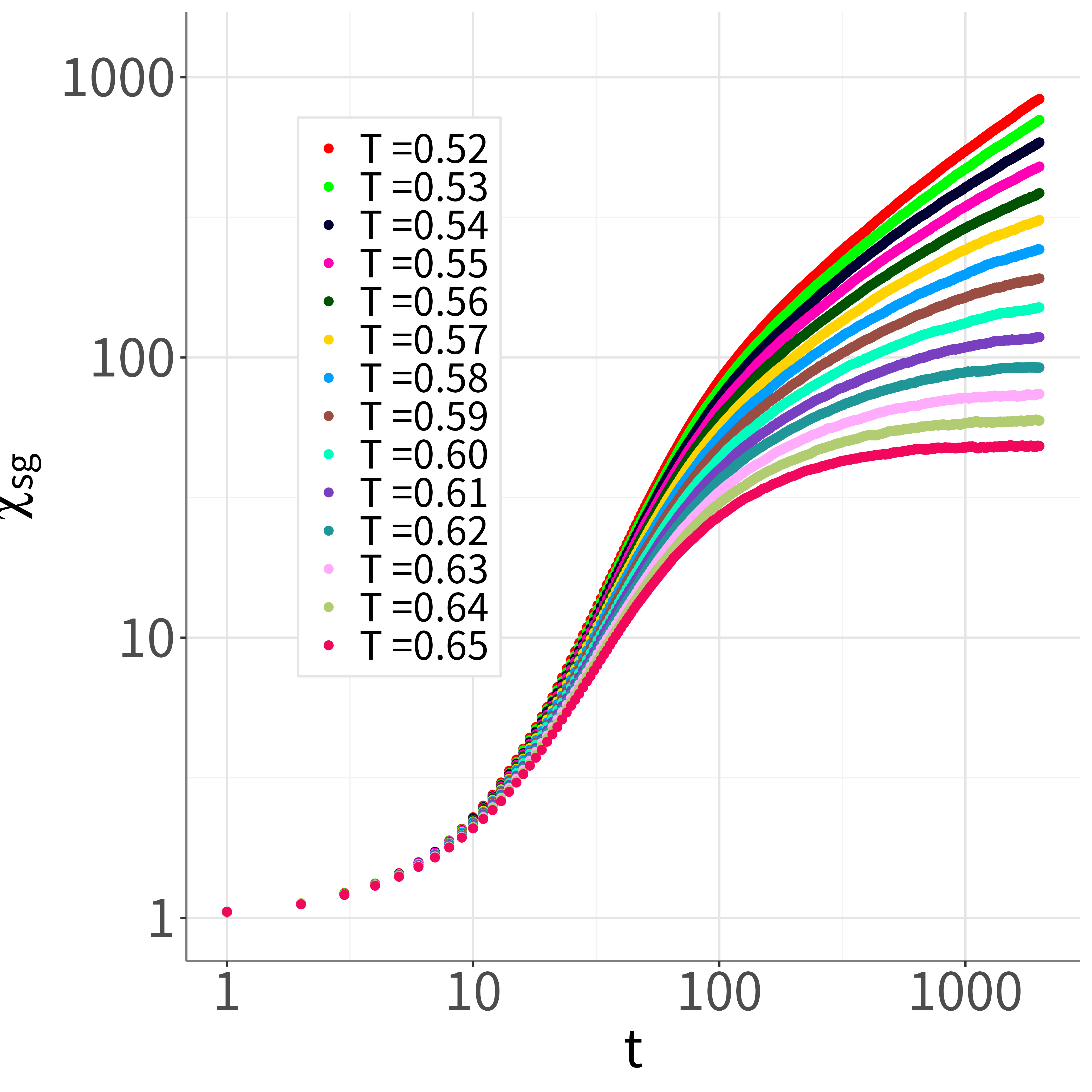}
                \centerline{(a)}
                \includegraphics[width=6cm]{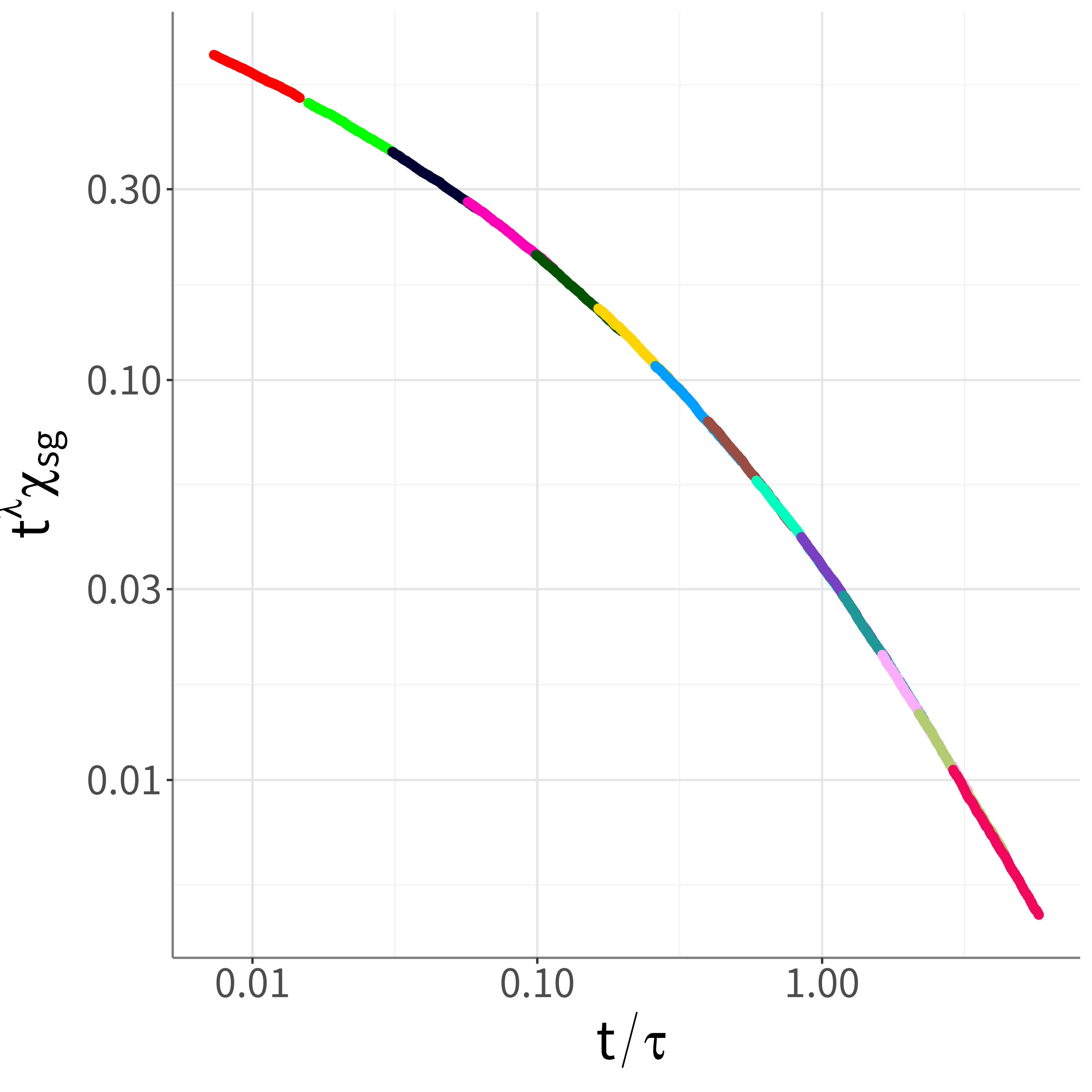}
                \centerline{(b)}
        \caption{
        \label{figure:d0_plot}
            (a) Relaxation of $\chi_{\sg}$ in the vicinity of $T_{\sg}$ for $1/D = \infty$ with $L = 151$, covering the range $0.52 \le T \le 0.65$.
            (b) Corresponding scaling plot according to \cref{eq:sus_dynamical_scaling}.
            The initial relaxation data for $t < 450$ are excluded from the scaling analysis.
            The estimated parameters are $T_{\sg} = 0.4397(13)$, $\gamma/(z\nu) = 0.8240(1)$, and $z\nu = 5.545(33)$.}
        \end{figure}

    To demonstrate our analysis procedure, we first present the detailed results for $1/D = \infty$ as a representative case.
    \Cref{figure:d0_plot}(a) shows the relaxation data for the SG susceptibility $\chi_{\sg}$ for $0.52 \le T \le 0.65$ with $L=151$.
    Performing dynamical scaling plot according to \cref{eq:sus_dynamical_scaling}, we obtained the data collapse shown in \cref{figure:d0_plot}(b), where we obtained fitting parameters as $T_{\sg} = 0.4397(13)$, $\gamma/(z\nu) = 0.8240(1)$, and $z\nu = 5.545(33)$.

\begin{figure}[htbp]
            \centering
                \includegraphics[width=6cm]{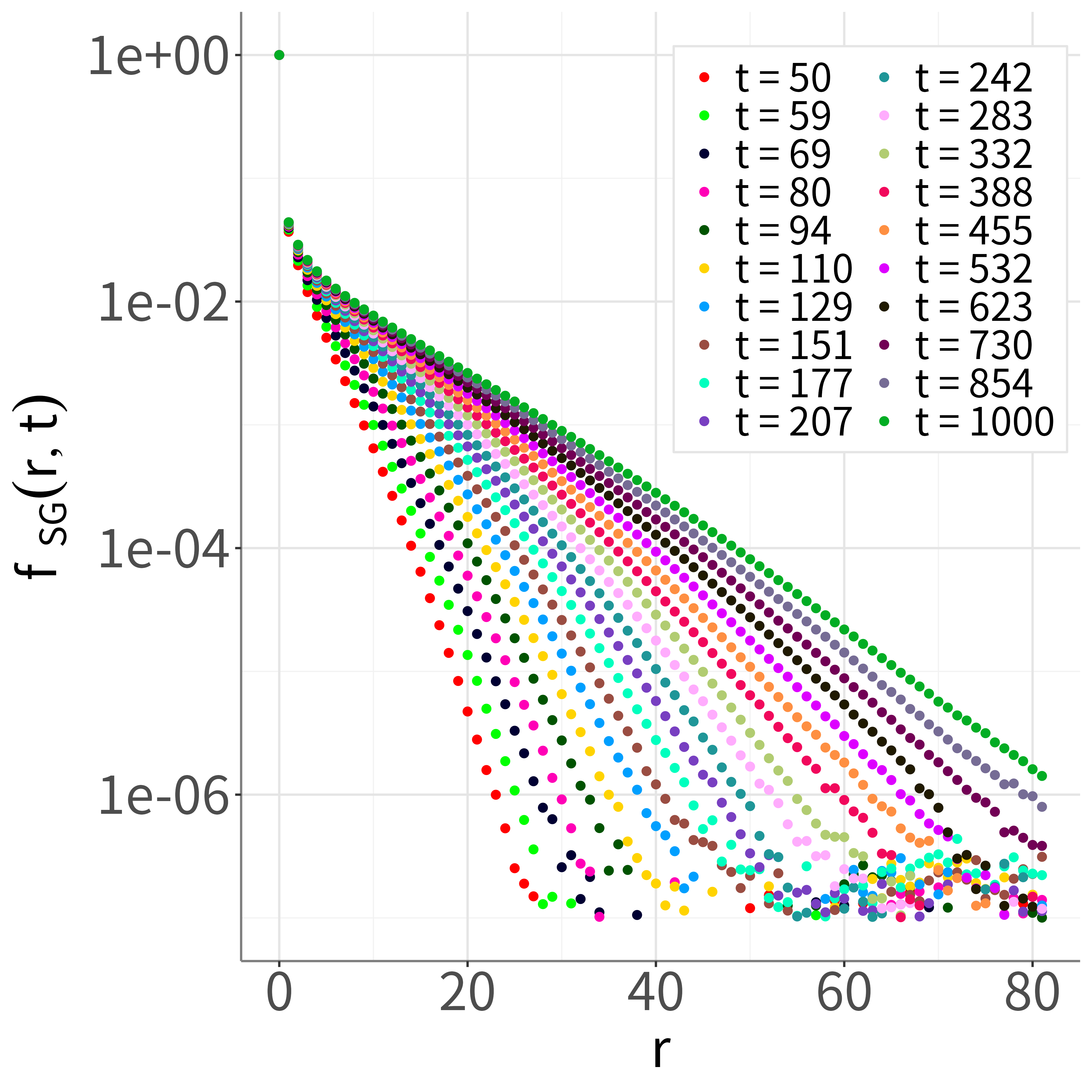}
                \centerline{(a)}
                \includegraphics[width=6cm]{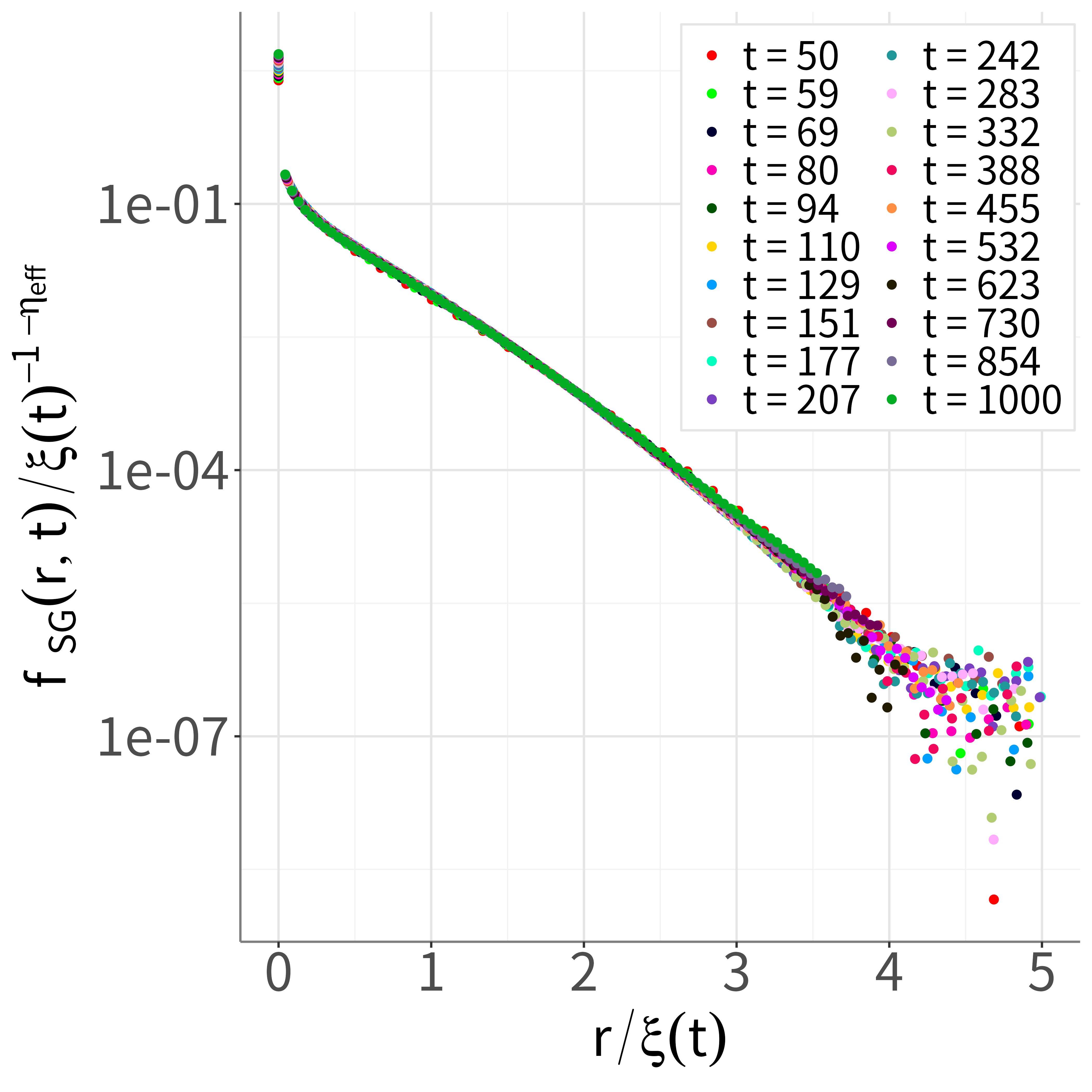}
                \centerline{(b)}
            \caption{\label{figure:sg_correlation}
            (a) SG correlation function $f_{\sg}(r, t)$ as a function of $t$ at the SG transition temperature $T = 0.4397$ obtained above for $1/D=\infty$ with $L = 251$.
            The data points are sampled at logarithmic time intervals within the range $1 \le t \le 1000$.
            (b) Corresponding scaling collapse according to \cref{eq:correlation_function_scaling}.
            The scaling analysis is performed for the spatial range $0 \le r \le L/3$.
}
        \end{figure}

\begin{figure}[htbp]
            \centering
                \includegraphics[width=5cm]{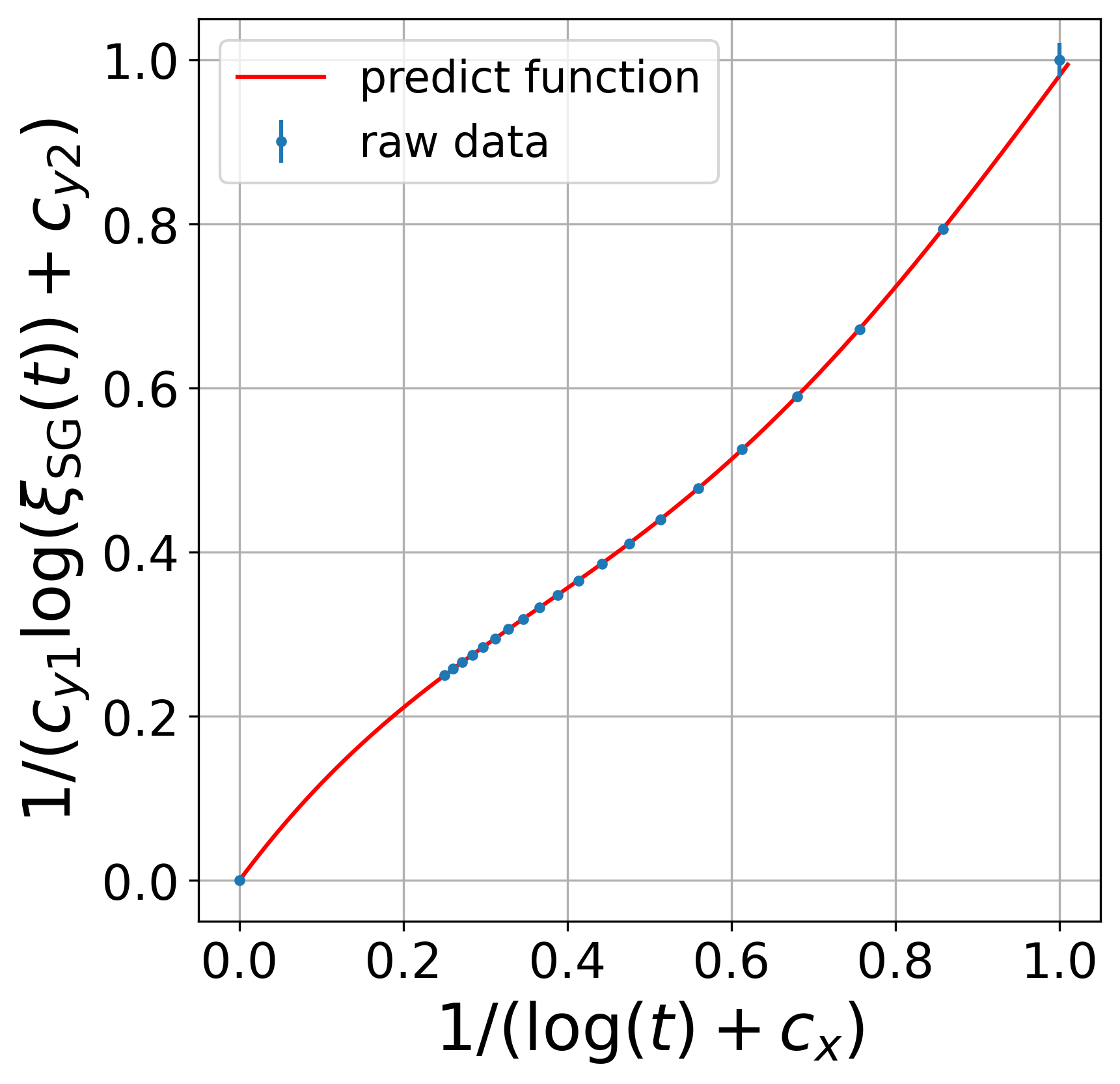}
                \centerline{(a)}
                \includegraphics[width=5cm]{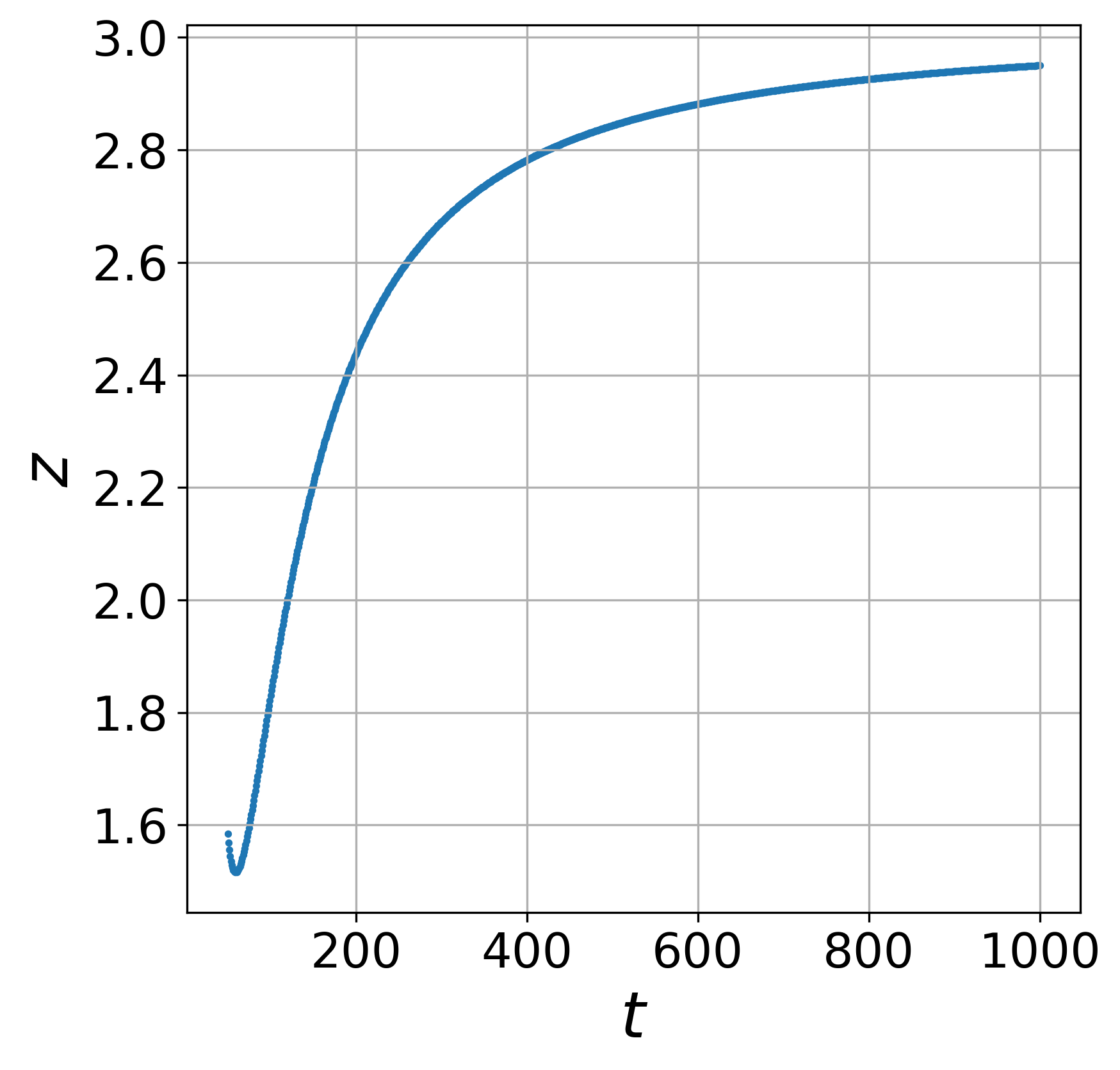}
                \centerline{(b)}
        \caption{\label{figure:xisg}
            (a) Estimated SG correlation length $\xi_\sg(t)$ for several values of $t$ plotted on the scale defined in \cref{eq:converted-data}.
            As in \cref{figure:regression_local}, the horizontal axis represents the inverse of the logarithm of $t$, and the vertical axis represents the inverse of the logarithm of $\xi_\sg(t)$.
            The orange curves denote the regressions performed using data sampled at equal logarithmic intervals, along with the data point (0,0) corresponding to the thermodynamic limit.
            The simulation data (black dots) align well with the regression curves, demonstrating that the fitted curves accurately reproduce the raw data.
            (b) Local exponent $z(t)$ defined in \cref{eq:def_z} obtained from the derivatives of the regression curves in (a).
        }
        \end{figure}

    \Cref{figure:sg_correlation}(a) shows the data for the SG correlation function at $T=T_{\sg}$ obtained above.
    The corresponding scaling plot using \cref{eq:correlation_function_scaling} is given by \cref{figure:sg_correlation}(b), where the scaling parameters, $\xi_\sg (t)$ for each $t$ and $\eta_{\mathrm{eff}}$, are obtained as $\eta_{\mathrm{eff}} = -0.495(13)$ and the values of $\xi_\sg (t)$ shown in \cref{figure:xisg}(a).
    Using the scheme in \cref{subsubsec:gpr}, in \cref{figure:xisg}(a), we plotted the data as in \cref{figure:regression_local} together with the regression curve.
    Then, we estimate the local exponent $z(t)$ and plot it in \cref{figure:xisg}(b), which gives the asymptotic value $z=3.01(1)$ by extrapolation.
    Using these results and the scaling parameters $z\nu$ and $\gamma / (z\nu)$, the values $\nu = 1.84(1)$ and $\eta = -0.48(2)$ are obtained.

\begin{figure}[htbp]
            \centering
            \includegraphics[width=6cm]{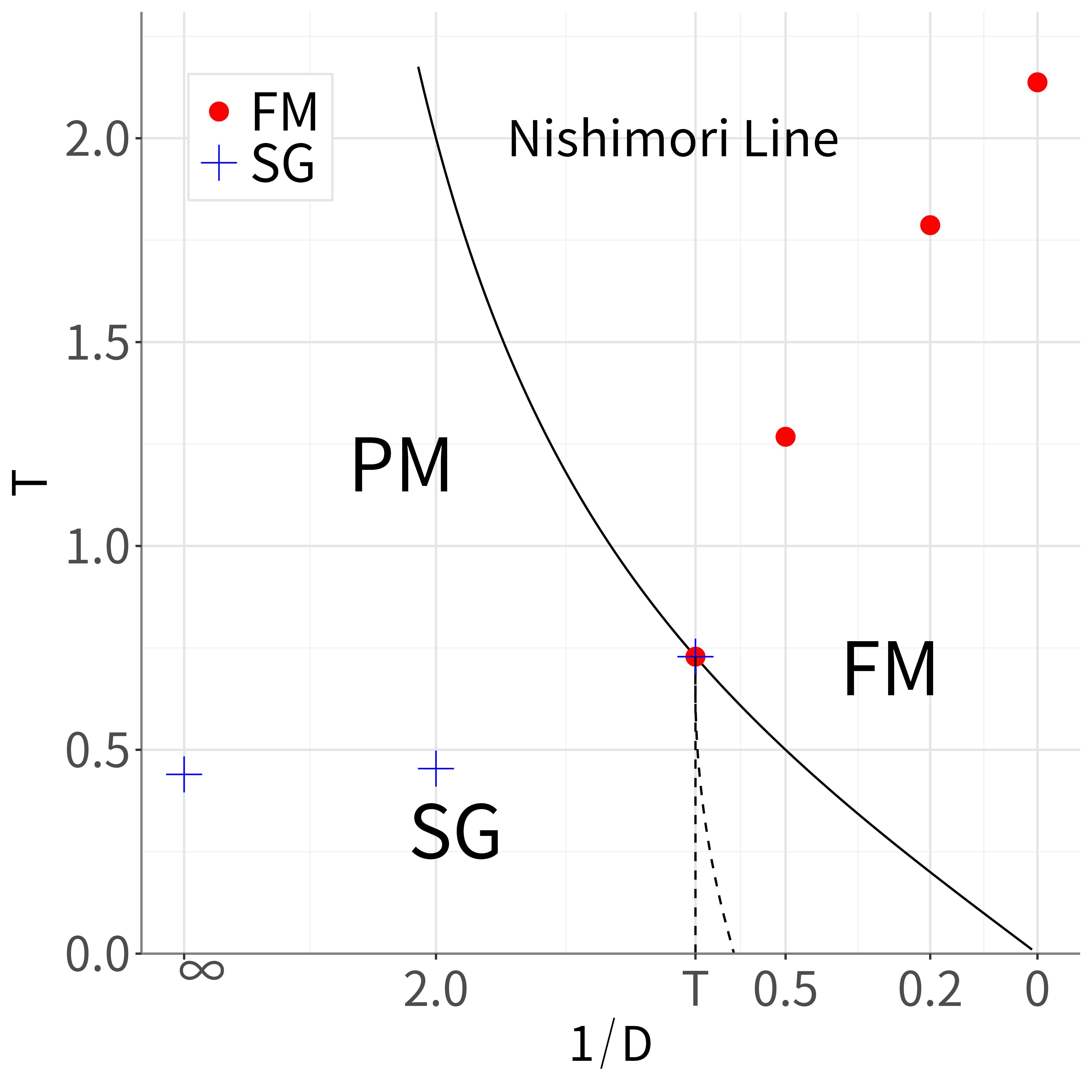}
            \caption{\label{figure:souzu}
            Resulting phase diagram of the 3D $SU(2)$ GG model.
            The horizontal axis is scaled by the $\arctan$ function.
            Although the phase boundary of the FM-SG transition is not explicitly determined in this study, gauge theory dictates that the boundary should be either vertical or reentrant.}
        \end{figure}

        Following the same procedure, we also evaluated $T_{\sg}$ and the critical exponents for $1/D = 2.0$.
        \Cref{table:sg_transition} summarizes the results for both cases.
        As seen in \cref{table:sg_transition}, $T_{\sg}$ decreases as $1/D$ increases.
        In contrast, $\nu$ and $\eta$ are consistent within the error bars, suggesting that these cases belong to the same universality class.
        Additionally, $z$ tends to increase with $1/D$, similar to the behavior observed in the FM-PM transition.
        Based on the critical temperatures obtained from these analyses, we construct the overall phase diagram of the $SU(2)$ GG model, as shown in \cref{figure:souzu}.

        \begin{table}[htpb]
            \caption{Results for SG critical exponents.
            For $1/D = 2.0$ and $\infty$, the critical exponents $\nu$ and $\eta$ are consistent within statistical errors. This suggests that these cases belong to the same universality class.}
        \label{table:sg_transition}
        \vspace{2mm}
        \centering
        \begin{tabular}{c|llll}
        \hline
        \hline
        $1/D$ & $T_{\sg}$ & $\nu$ & $\eta$ & $z$\\
        \hline
        2.0 & 0.450(2) & 1.81(4) & -0.40(6) & 2.93(6) \\
        $\infty$ & 0.440(1) & 1.84(1) & -0.48(2) & 3.01(1)\\
        \hline
        \hline
        \end{tabular}
        \end{table}

\section{Discussion}
    \label{sec:discussion}
    In this study, we numerically analyzed the phase transitions of the 3D $SU(2)$ gauge glass model and investigated the influence of quenched disorder on the critical universality class.
    We performed high-precision evaluations of the transition temperature and critical exponents using dynamical scaling analysis and relaxation of fluctuations analysis within the framework of the non-equilibrium relaxation method.
    In both analyses, a machine learning technique, Gaussian process regression with the kernel method, was applied to ensure the precision and reliability of the results.

    As shown in \cref{table:o4_comparison}, the results for the $O(4)$ model are consistent with those of previous studies.
    The static critical exponents at $1/D=0.2$ and $0.5$ in \cref{table:critical_exponent} are also consistent with those at $1/D=0$, and the obtained exponents agree with those of the 3D $O(4)$ universality class.
    These results indicate that the $O(4)$ model is robust against weak quenched disorder, with its universality class remaining unchanged.
    Furthermore, these findings are consistent with the Harris criterion.

    At the MCP, however, the critical exponents differ significantly from those at other points on the phase boundary, indicating that the system belongs to a distinct universality class.
    Specifically, the critical exponent $\nu = 0.987(1)$ obtained at the MCP exhibits a statistically significant deviation from the superuniversality hypothesis of $\nu = 1$ proposed in Ref.~\onlinecite{Delfino_2025}.
    It is noted that similar deviations have also been reported in previous studies for other GG models.
    For instance,
    $\nu = 0.90(2)$~\cite{doi:10.7566/JPSJ.92.074003} for the 3D Ising SG model, and $\nu = 1.07(3)$~\cite{PhysRevB.83.094203} for the 3D XY GG model.
    While this widespread discrepancy could be interpreted as evidence that the superuniversality hypothesis has failed, it is also possible that it stems from significant scaling corrections near the MCP.
    In fact, in the 3D Ising SG model, there are also studies that report values consistent with superuniversality --- such as $\nu = 0.98(5)$~\cite{PhysRevB.76.184202} --- in addition to the examples mentioned above.
    To reach a definitive conclusion, further investigation that explicitly accounts for these strong corrections will likely be necessary.

    For the SG transition, we confirmed the existence of a finite-temperature SG phase in the $SU(2)$ GG model.
    As shown in \cref{table:sg_transition}, we found that this model indicates a universality class that is independent of the degrees of disorder.
    The critical exponents obtained in the pure SG limit ($1/D = \infty$) are $\nu = 1.84(1)$ and $\eta = -0.48(2)$.
    The non-equilibrium relaxation analysis employed in this study is a well-established method for analyzing SG transitions and critical phenomena, and is expected to serve as a useful reference for future extensions and applications especially for continuous-spin models.
    Since only two types of disorder were analyzed, further verification across a wider range of disorder levels is necessary to draw conclusions regarding critical universality.
    In particular, SG transitions in the vicinity of the MCP may be strongly influenced by fluctuations, necessitating a more careful analysis; however, we leave this as a topic for future research.
    The prediction that SG transitions constitute a single universality class independent of disorder is widely accepted as a matter of common sense; however, due to the difficulty of analyzing SG systems, there are few examples of systematic verification. The $SU(2)$ GG model, confirmed following the Ising SG model, is expected to serve as a catalyst for similar analyses of more realistic XY SG and Heisenberg SG systems.

\section*{Acknowledgments}
This work was supported by JST SPRING, Grant Number JPMJSP2131.
The authors are also grateful to the Supercomputer Center at the Institute for Solid State Physics, University of Tokyo, for the use of their facilities.

\bibliographystyle{jpsj_mod}
\bibliography{statphys29_takada.bib}

@string{JPF = "J. Phys. F: Met. Phys."}

@string{JSTAT = "J. Stat. Mech."}

@string{NATURE = "Nature"}

@string{MTAC = "Math. Tables Aids Comput."}

@string{JMP = "J. Math. Phys."}

@string{JPAG = "J. Phys. A: Math. Gen."}

@string{JPAMT = "J. Phys. A: Math. Theor."}

@string{JPSJ="J. Phys. Soc. Jpn."}

@article{10.1143/PTP.66.1169,
    author = {H.Nishimori},
    title = {Internal Energy, Specific Heat and Correlation Function of the Bond-Random Ising Model},
    journal = {Prog. Theor. Phys.},
    volume = {66},
    number = {4},
    pages = {1169-1181},
    year = {1981},
    month = {10},
    issn = {0033-068X},
    doi = {10.1143/PTP.66.1169},
    url = {https://doi.org/10.1143/PTP.66.1169},
    eprint = {https://academic.oup.com/ptp/article-pdf/66/4/1169/5265369/66-4-1169.pdf},
}

@article{PhysRevE.94.043312,
  title = {Improved dynamical scaling analysis using the kernel method for nonequilibrium relaxation},
  author = {Echinaka, Y and Ozeki, Y},
  journal = {Phys. Rev. E},
  volume = {94},
  issue = {4},
  pages = {043312},
  numpages = {9},
  year = {2016},
  month = {Oct},
  publisher = {American Physical Society},
  doi = {10.1103/PhysRevE.94.043312},
  url = {https://link.aps.org/doi/10.1103/PhysRevE.94.043312}
}

@article{A_B_Harris_1974,
doi = {10.1088/0022-3719/7/9/009},
url = {https://dx.doi.org/10.1088/0022-3719/7/9/009},
year = {1974},
month = {may},
publisher = {},
volume = {7},
number = {9},
pages = {1671},
author = {A B Harris},
title = {Effect of random defects on the critical behaviour of Ising models},
journal = {J. Phys. C: Solid State Phys.}
}

@article{ozeki2007nonequilibrium,
  title={Nonequilibrium relaxation method},
  author={Y. Ozeki and N. Ito},
  journal= JPAMT,
  volume={40},
  number={31},
  pages={R149},
  year={2007},
  publisher={IOP Publishing}
}

@article{doi:10.7566/JPSJ.93.114001,
author = {Y. Osada and Y. Ozeki},
title = {Improvement of Analysis for Relaxation of Fluctuations by Using Gaussian Process Regression and Extrapolation Method},
journal = JPSJ,
volume = {93},
number = {11},
pages = {114001},
year = {2024},
doi = {10.7566/JPSJ.93.114001},

URL = {https://doi.org/10.7566/JPSJ.93.114001},
eprint = {https://doi.org/10.7566/JPSJ.93.114001}
}

@article{Ozeki1993,
  doi = {10.1088/0305-4470/26/14/009},
  year = {1993},
  url = {https://dx.doi.org/10.1088/0305-4470/26/14/009},
  volume = {26},
  number = {14},
  pages = {3399},
  author = {Y. Ozeki and H. Nishimori},
  title = {Phase diagram of gauge glasses},
  journal = JPAG
}

@article{PhysRevD.51.2404,
  title = {Critical exponents of a three-dimensional O(4) spin model},
  author = {Kanaya, K. and Kaya, S.},
  journal = {Phys. Rev. D},
  volume = {51},
  issue = {5},
  pages = {2404--2410},
  numpages = {0},
  year = {1995},
  month = {Mar},
  publisher = {American Physical Society},
  doi = {10.1103/PhysRevD.51.2404},
  url = {https://link.aps.org/doi/10.1103/PhysRevD.51.2404}
}

@article{Kos2015,
  author    = {Filip Kos and David Poland and David Simmons-Duffin and Alessandro Vichi},
  title     = {Bootstrapping the O(N) archipelago},
  journal   = {J. High Energy Phys.},
  volume    = {2015},
  number    = {11},
  pages     = {106},
  year      = {2015},
  month     = {Nov},
  doi       = {10.1007/JHEP11(2015)106},
  url       = {https://doi.org/10.1007/JHEP11(2015)106}
}

@article{PhysRevE.110.064108,
  title = {Anomalous distribution of magnetization in an Ising spin glass with correlated disorder},
  author = {Nishimori, Hidetoshi},
  journal = {Phys. Rev. E},
  volume = {110},
  issue = {6},
  pages = {064108},
  numpages = {9},
  year = {2024},
  month = {Dec},
  publisher = {American Physical Society},
  doi = {10.1103/PhysRevE.110.064108},
  url = {https://link.aps.org/doi/10.1103/PhysRevE.110.064108}
}

@article{Delfino_2025,
  doi = {10.1088/1742-5468/adc4ce},
  year = {2025},
  volume = {2025},
  number = {4},
  pages = {043203},
  author = {Delfino, Gesualdo},
  title = {Critical exponents at the Nishimori point},
  journal = JSTAT,
}

@book{Mydosh1993,
  author    = {Mydosh, J. A.},
  title     = {Spin Glasses: An Experimental Introduction},
  publisher = {CRC Press},
  year      = {1993},
  edition   = {1},
  address   = {London},
  doi       = {10.1201/9781482295191},
  url       = {https://www.taylorfrancis.com/books/mono/10.1201/9781482295191/spin-glasses-mydosh}
}

@article{PhysRevB.33.6533,
  title = {Quenched disorder in Josephson-junction arrays in a transverse magnetic field},
  author = {Granato, Enzo and Kosterlitz, J. M.},
  journal = {Phys. Rev. B},
  volume = {33},
  issue = {9},
  pages = {6533--6536},
  numpages = {0},
  year = {1986},
  month = {May},
  publisher = {American Physical Society},
  doi = {10.1103/PhysRevB.33.6533},
  url = {https://link.aps.org/doi/10.1103/PhysRevB.33.6533}
}

@article{PhysRevB.42.1059,
  title = {Possible vortex-glass transition in a model random superconductor},
  author = {Huse, David A. and Seung, H. Sebastian},
  journal = {Phys. Rev. B},
  volume = {42},
  issue = {1},
  pages = {1059--1061},
  numpages = {0},
  year = {1990},
  month = {Jul},
  publisher = {American Physical Society},
  doi = {10.1103/PhysRevB.42.1059},
  url = {https://link.aps.org/doi/10.1103/PhysRevB.42.1059}
}

@article{Sourlas1989,
  author  = {Sourlas, Nicolas},
  title   = {Spin-glass models as error-correcting codes},
  journal = {Nature},
  year    = {1989},
  volume  = {339},
  issue   = {6227},
  pages   = {693--695},
  month   = {Jun},
  doi     = {10.1038/339693a0},
  url     = {https://doi.org/10.1038/339693a0},
  issn    = {1476-4687}
}

@article{F_Barahona_1982,
  doi = {10.1088/0305-4470/15/10/028},
  url = {https://doi.org/10.1088/0305-4470/15/10/028},
  year = {1982},
  volume = {15},
  number = {10},
  pages = {3241},
  author = {F Barahona},
  title = {On the computational complexity of Ising spin glass models},
  journal = JPAG
}

@article{PhysRevE.104.064101,
  title = {Precision calculation of universal amplitude ratios in O($N$) universality classes: Derivative expansion results at order $\mathcal{O}({\ensuremath{\partial}}^{4})$},
  author = {De Polsi, Gonzalo and Hern\'andez-Chifflet, Guzm\'an and Wschebor, Nicol\'as},
  journal = {Phys. Rev. E},
  volume = {104},
  issue = {6},
  pages = {064101},
  numpages = {22},
  year = {2021},
  month = {Dec},
  publisher = {American Physical Society},
  doi = {10.1103/PhysRevE.104.064101},
  url = {https://link.aps.org/doi/10.1103/PhysRevE.104.064101}
}

@article{PhysRevB.105.054428,
  title = {Three-dimensional $O(N)$-invariant ${\ensuremath{\phi}}^{4}$ models at criticality for $N\ensuremath{\ge}4$},
  author = {Hasenbusch, Martin},
  journal = {Phys. Rev. B},
  volume = {105},
  issue = {5},
  pages = {054428},
  numpages = {16},
  year = {2022},
  month = {Feb},
  publisher = {American Physical Society},
  doi = {10.1103/PhysRevB.105.054428},
  url = {https://link.aps.org/doi/10.1103/PhysRevB.105.054428}
}

@article{PhysRevD.29.338,
  title = {Remarks on the chiral phase transition in chromodynamics},
  author = {Pisarski, Robert D. and Wilczek, Frank},
  journal = {Phys. Rev. D},
  volume = {29},
  issue = {2},
  pages = {338--341},
  numpages = {0},
  year = {1984},
  month = {Jan},
  publisher = {American Physical Society},
  doi = {10.1103/PhysRevD.29.338},
  url = {https://link.aps.org/doi/10.1103/PhysRevD.29.338}
}

@article{PhysRevB.72.014462,
  title = {Dynamical scaling in Ising and vector spin glasses},
  author = {Katzgraber, Helmut G. and Campbell, I. A.},
  journal = {Phys. Rev. B},
  volume = {72},
  issue = {1},
  pages = {014462},
  numpages = {13},
  year = {2005},
  month = {Jul},
  publisher = {American Physical Society},
  doi = {10.1103/PhysRevB.72.014462},
  url = {https://link.aps.org/doi/10.1103/PhysRevB.72.014462}
}

@article{nakamura2006nonequilibriumdynamicexponentspinglass,
  title={Nonequilibrium dynamic exponent and spin-glass transitions},
  url={https://arxiv.org/abs/cond-mat/0603062},
  author={Tota Nakamura},
  year={2006},
  journal={arXiv:cond-mat/0603062}
}

@article{Nakamura_2019,
   DOI={10.1103/physreve.99.023301},
   url={http://dx.doi.org/10.1103/PhysRevE.99.023301},
   volume={99},
   number={2},
   pages={023301},
   journal={Phys. Rev. E},
   author={Nakamura, Tota},
   year={2019}
}

@article{doi:10.7566/JPSJ.92.074003,
author = {Terasawa ,Yusuke and Ozeki ,Yukiyasu},
title = {Dynamical Scaling Analysis for ±J Ising Model in Three Dimensions},
journal = {J. Phys. Soc. Jpn.},
volume = {92},
number = {7},
pages = {074003},
year = {2023},
doi = {10.7566/JPSJ.92.074003},
URL = {https://doi.org/10.7566/JPSJ.92.074003},
}

@article{PhysRevB.83.094203,
  title = {Temperature-disorder phase diagram of a three-dimensional gauge-glass model},
  author = {Alba, Vincenzo and Vicari, Ettore},
  journal = {Phys. Rev. B},
  volume = {83},
  issue = {9},
  pages = {094203},
  numpages = {11},
  year = {2011},
  month = {Mar},
  publisher = {American Physical Society},
  doi = {10.1103/PhysRevB.83.094203},
  url = {https://link.aps.org/doi/10.1103/PhysRevB.83.094203}
}

@ARTICLE{1975JPhF5965E,
       author = {{Edwards}, S.~F. and {Anderson}, P.~W.},
        title = "{Theory of spin glasses}",
      journal = JPF,
         year = {1975},
       volume = {5},
        pages = {965-974},
          doi = {10.1088/0305-4608/5/5/017}
}

@article{RevModPhys.58.801,
  title = {Spin glasses: Experimental facts, theoretical concepts, and open questions},
  author = {Binder, K. and Young, A. P.},
  journal = {Rev. Mod. Phys.},
  volume = {58},
  issue = {4},
  pages = {801--976},
  numpages = {0},
  year = {1986},
  month = {Oct},
  publisher = {American Physical Society},
  doi = {10.1103/RevModPhys.58.801},
  url = {https://link.aps.org/doi/10.1103/RevModPhys.58.801}
}

@book{Charbonneau2023,
  title     = {Spin Glass Theory and Far Beyond: Replica Symmetry Breaking after 40 Years},
  editor    = {Charbonneau, Patrick and Marinari, Enzo and M{\'e}zard, Marc and Parisi, Giorgio and Ricci-Tersenghi, Federico and Sicuro, Gabriele and Zamponi, Francesco},
  year      = {2023},
  publisher = {World Scientific},
  address   = {Singapore},
  doi       = {10.1142/13341},
  isbn      = {978-981-127-391-9}
}

@article{Nakamura_2016,
   title={From measurements to inferences of physical quantities in numerical simulations},
   volume={93},
   ISSN={2470-0053},
   url={http://dx.doi.org/10.1103/PhysRevE.93.011301},
   DOI={10.1103/physreve.93.011301},
   number={1},
   pages = {011301},
   journal={Phys. Rev. E},
   publisher={American Physical Society (APS)},
   author={Nakamura, Tota},
   year={2016},
   month=Jan
}

@article{Yukiyasu_Ozeki_2003,
  doi = {10.1088/0305-4470/36/11/303},
  url = {https://doi.org/10.1088/0305-4470/36/11/303},
  year = {2003},
  volume = {36},
  number = {11},
  pages = {2673},
  author = {Yukiyasu Ozeki},
  title = {Dynamical gauge theory for the XY gauge glass model},
  journal = JPAG
}

@article{PhysRevE.111.044109,
  title = {Instability of the ferromagnetic phase under random fields in an Ising spin glass with correlated disorder},
  author = {Nishimori, Hidetoshi},
  journal = {Phys. Rev. E},
  volume = {111},
  issue = {4},
  pages = {044109},
  numpages = {9},
  year = {2025},
  month = {Apr},
  publisher = {American Physical Society},
  doi = {10.1103/PhysRevE.111.044109},
  url = {https://link.aps.org/doi/10.1103/PhysRevE.111.044109}
}

@article{qp1w-qcbs,
  title = {Temperature chaos as a logical consequence of the reentrant transition in spin glasses},
  author = {Nishimori, Hidetoshi and Ohzeki, Masayuki and Okuyama, Manaka},
  journal = {Phys. Rev. E},
  volume = {112},
  issue = {4},
  pages = {044140},
  numpages = {12},
  year = {2025},
  month = {Oct},
  publisher = {American Physical Society},
  doi = {10.1103/qp1w-qcbs},
  url = {https://link.aps.org/doi/10.1103/qp1w-qcbs}
}

@article{PhysRevLett.35.1792,
  title = {Solvable Model of a Spin-Glass},
  author = {Sherrington, David and Kirkpatrick, Scott},
  journal = {Phys. Rev. Lett.},
  volume = {35},
  issue = {26},
  pages = {1792--1796},
  numpages = {0},
  year = {1975},
  month = {Dec},
  publisher = {American Physical Society},
  doi = {10.1103/PhysRevLett.35.1792},
  url = {https://link.aps.org/doi/10.1103/PhysRevLett.35.1792}
}

@article{terasawa2026neural,
  title={Dynamical scaling method improved by a deep learning approach},
  author = {Y. Terasawa and Y. Ozeki},
    journal = JPSJ,
    volume = {95},
    pages = {064004},
    year = {2026}
}

@article{terasawa2026sg,
  title={Dynamical scaling study for the estimation of dynamical exponent   of three-dimensional XY spin glass model},
  url={https://arxiv.org/abs/2503.06386},
  author = {Y. Terasawa and Y. Ozeki},
  year={2026},
  journal={arXiv:cond-mat/2503.06386}
}

@article{PhysRevB.76.184202,
  title = {Magnetic-glassy multicritical behavior of the three-dimensional $\ifmmode\pm\else\textpm\fi{}J$ Ising model},
  author = {Hasenbusch, Martin and Toldin, Francesco Parisen and Pelissetto, Andrea and Vicari, Ettore},
  journal = {Phys. Rev. B},
  volume = {76},
  issue = {18},
  pages = {184202},
  numpages = {7},
  year = {2007},
  month = {Nov},
  publisher = {American Physical Society},
  doi = {10.1103/PhysRevB.76.184202},
  url = {https://link.aps.org/doi/10.1103/PhysRevB.76.184202}
}

@article{PhysRevB.70.184417,
  title = {Nonvanishing spin-glass transition temperature of the $\ifmmode\pm\else\textpm\fi{}J$ $XY$ model in three dimensions},
  author = {Yamamoto, Takeo and Sugashima, Takeshi and Nakamura, Tota},
  journal = {Phys. Rev. B},
  volume = {70},
  issue = {18},
  pages = {184417},
  numpages = {10},
  year = {2004},
  month = {Nov},
  publisher = {American Physical Society},
  doi = {10.1103/PhysRevB.70.184417},
  url = {https://link.aps.org/doi/10.1103/PhysRevB.70.184417}
}

@article{PhysRevE.79.041138,
  title = {Effects of discreteness on gauge glass models in two and three dimensions},
  author = {Yotsuyanagi, Satoshi and Suemitsu, Yusuke and Ozeki, Yukiyasu},
  journal = {Phys. Rev. E},
  volume = {79},
  issue = {4},
  pages = {041138},
  numpages = {9},
  year = {2009},
  month = {Apr},
  publisher = {American Physical Society},
  doi = {10.1103/PhysRevE.79.041138},
  url = {https://link.aps.org/doi/10.1103/PhysRevE.79.041138}
}

@article{PhysRevB.82.014427,
  title = {Nonequilibrium dynamic correlation-length scaling method},
  author = {Nakamura, Tota},
  journal = {Phys. Rev. B},
  volume = {82},
  issue = {1},
  pages = {014427},
  numpages = {7},
  year = {2010},
  month = {Jul},
  publisher = {American Physical Society},
  doi = {10.1103/PhysRevB.82.014427},
  url = {https://link.aps.org/doi/10.1103/PhysRevB.82.014427}
}

@article{wynn1956device,
 ISSN = {08916837},
 URL = {http://www.jstor.org/stable/2002183},
 author = {P. Wynn},
 journal = MTAC,
 number = {54},
 pages = {91--96},
 publisher = {American Mathematical Society},
 title = {On a Device for Computing the em(Sn) Transformation},
 urldate = {2024-09-02},
 volume = {10},
 year = {1956}
}

@book{sidi2003practical,
  place={Cambridge},
  title={Practical Extrapolation Methods: Theory and Applications},
  publisher={Cambridge University Press},
  author={Sidi, Avram},
  year={2003},
  url={https://www.cambridge.org/core/books/practical-extrapolation-methods/21A93C2B0793CF09B2F3ABEF78F3F9B9}
}

@book{brezinski2020extrapolation,
  place={Cham},
  title={Extrapolation and Rational Approximation},
  publisher={Springer},
  author={Brezinski, Claude and Redivo-Zaglia, Michela},
  year={2020},
  url={https://link.springer.com/book/10.1007/978-3-030-58418-4}
}

@article{shanks1955non,
  title={Non-linear transformations of divergent and slowly convergent sequences},
  author={Shanks, Daniel},
  journal=JMP,
  volume={34},
  number={1-4},
  pages={1--42},
  year={1955},
  publisher={Wiley Online Library},
  url={https://onlinelibrary.wiley.com/doi/abs/10.1002/sapm19553411}
}
\end{document}